\documentclass[12pt,a4paper]{article}
\usepackage{a4wide}
\usepackage{amsmath}
\usepackage{amssymb}
\usepackage{epsfig}
\usepackage{subfigure}
\usepackage{exscale}
\usepackage{float}
\usepackage{bbm}
\usepackage[numbers,sort&compress]{natbib}
\usepackage{pst-plot, pstricks,pst-math}
\usepackage{fancybox,amssymb,color}
\usepackage{graphicx}
\usepackage{pstricks, color, graphicx, epsfig, psfrag}
\usepackage{amsfonts,amsmath,amssymb,slashed}
\usepackage{dsfont}
\usepackage{bbm,bm}
\usepackage{fancyhdr, a4wide}
\usepackage[english]{babel}
\usepackage{subfigure}

\makeatletter
\@addtoreset{equation}{section}
\makeatother
\renewcommand{\theequation}{\thesection.\arabic{equation}}

\title{Thermodynamics of O(N) Antiferromagnets in 2+1 Dimensions}

\author{Christoph P.\ Hofmann$^a$ \\ \\
\normalsize {$^a$ Facultad de Ciencias, Universidad de Colima} \\
\vspace{0.3cm}
\normalsize {Bernal D\'iaz del Castillo 340, Colima C.P.\ 28045, Mexico} \\}

\begin{document}
\maketitle

\begin{abstract} \normalsize

Within the framework of effective Lagrangians we calculate the free energy
density for an O($N$) antiferromagnet in 2+1 dimensions up to three-loop
order in the perturbative expansion and derive the low-temperature series for
various thermodynamic quantities. In particular, we show that the magnon-magnon
interaction in the O(3) antiferromagnet in $d$=2+1 -- the O(3)-invariant
quantum Heisenberg antiferromagnet on a square or a honeycomb lattice -- is
very weak and repulsive and manifests itself through a term proportional to
five powers of the temperature in the free energy density. Remarkably, the
corresponding coefficient is fully determined by the leading-order effective
Lagrangian ${\cal L}^2_{eff}$ and does not involve any higher order effective
constants from ${\cal L}^4_{eff}$ related to the anisotropies of the lattice
-- the symmetries are thus very restrictive in $d$=2+1. We also compare our
results that apply to O($N$) antiferromagnets in 2+1 dimensions with the those
for O($N$) antiferromagnets in 3+1 dimensions. The present work demonstrates
the efficiency of the fully systematic effective Lagrangian method in the
condensed matter domain, which clearly proves to be superior to spin-wave
theory. We would like to emphasize that the structure of the low-temperature
series derived in the present work is model-independent and universal as it
only relies on symmetry considerations. 

\end{abstract}


\maketitle

\section{Introduction}

\label{Intro}

With the present article we would like to further promote the effective
Lagrangian method in the condensed matter domain -- in particular, we would
like to demonstrate its efficiency in describing the thermodynamic properties
of systems exhibiting collective magnetic behavior. While the low-energy
properties of antiferromagnets in dimension $d$=3+1 have previously been
investigated within the effective Lagrangian framework
\cite{Hasenfratz Leutwyler,Leutwyler NRD,HofmannSpinWave,HofmannAF}, in the
present study we focus our attention on antiferromagnets in dimension $d$=2+1.
A thorough analysis of these condensed matter systems, using effective field
theory methods, was performed in
Refs.~\cite{Hasenfratz Leutwyler,Chakravarty Halperin Nelson,Neuberger Ziman,
Fisher,Hasenfratz Niedermayer}.
However, in the present paper, we go one step further in the perturbative
expansion, taking into account contributions to the free energy density up to
three-loop order. Our calculation applies to any system with a spontaneously
broken internal symmetry O($N$) $\to$ O($N$-1), provided that the
corresponding leading-order effective Lagrangian can be brought to a
Lorentz-invariant form. This system will be referred to as O($N$)
antiferromagnet. Throughout the paper, when talking about dimension, we always
refer to the space-time dimension $d=d_s+1$, where $d_s$ is the spatial
dimension.

The attentive reader may particularly wonder why the quantum Heisenberg
antiferromagnet,
\begin{equation}
\label{HeisenbergModel}
{\cal H} = -J \sum_{n.n.} {\vec S}_m \cdot {\vec S}_n \, , \qquad J=const. \, , 
\end{equation}
falls into the class of O(3) antiferromagnets, i.e., represents a system
described by a Lorentz-invariant leading-order effective Lagrangian. After
all, the lattice structure of a solid singles out preferred directions, such
that the effective Lagrangian in general is not even invariant under space
rotations. In the case of a cubic lattice, however, the anisotropy only shows
up at higher orders of the derivative expansion \cite{Hasenfratz Niedermayer}
-- the discrete symmetries of the three-dimensional cubic crystal thus imply
space rotation symmetry at leading order in the effective expansion. The same
is true for an antiferromagnet defined on a square or a honeycomb lattice,
which represents the system considered in this paper. Hence, the leading-order
effective Lagrangian describing the quantum Heisenberg antiferromagnet on a
cubic (d=3+1) or square/honeycomb (d=2+1) lattice is invariant under space
rotations and can be brought to a (pseudo-) Lorentz-invariant form
\cite{Leutwyler NRD}: Antiferromagnetic spin-wave excitations exhibit
relativistic kinematics, with the velocity of light replaced by the spin-wave
velocity. As we will explain in detail later on, the spatial anisotropies
which indeed start manifesting themselves at next-to-leading order in the
effective Lagrangian, will not affect at all the main result of the present
paper. Hence, a Lorentz-invariant framework, even at next-to-leading order of
the derivative expansion, is perfectly justified in our calculation.

Goldstone's theorem, which represents the basis of the systematic effective
Lagrangian method, states that, if the symmetry G = O($N$) of the Lagrangian is
spontaneously broken to H = O($N$-1), we must have $N$-1 Goldstone bosons in
the broken phase ($N \! \geq 2$). For $N$=3, these low-energy degrees of
freedom are identified with the two spin-wave excitations -- or the two
independent magnon particles -- in the spectrum of the O(3) antiferromagnet.

If the perturbations, which explicitly break the internal rotation symmetry
O($N$) of the Lagrangian, are small, the corresponding Goldstone excitations
remain light and dominate the low-energy behavior of the system. The effective
Lagrangian method thus also applies to antiferromagnets in weak external
fields. Goldstone's theorem guarantees that the Goldstone particles
interact only weakly at low energies such that a systematic perturbative
expansion in the momenta and the external fields can be performed. In the
present work we perturbatively evaluate the partition function in a power
series of the temperature, in order to obtain low-temperature theorems for
various quantities of physical interest.

We would like to emphasize that, from the effective Lagrangian perspective, the
analysis of the low-energy properties of the system is approached from a
unified and model-independent point of view, based on the spontaneously broken
symmetry of the system. The method applies to any system where the Goldstone
bosons are the only excitations without energy gap. The essential point is
that the properties of these low-energy degrees of freedom and their mutual
interaction are strongly constrained by the symmetries inherent in the
underlying theory, such as the Heisenberg Hamiltonian -- the specific nature
of the underlying theory or model, however, is not important. For general
pedagogic introductions to the effective Lagrangian technique see
Refs.~ \cite{Introductions}. Brief outlines of the method may be found in
Refs.~\cite{Outlines}. For specific applications to condensed matter systems
the reader may consult
Refs.~\cite{HofmannSpinWave,HofmannAF,HofmannFerro,Hasenfratz Niedermayer,
Roman Soto,Burgess,Uwe Construction,Uwe CEP,Uwe Application}.

Although our analysis is general, referring to any system which exhibits a
spontaneously broken symmetry O($N$) $\to$ O($N$-1) and a Lorentz-invariant
leading-order effective Lagrangian, our interest will primarily be devoted to
the special case $N$=3, which describes the O(3) Heisenberg
antiferromagnet in $d$=2+1 defined on a square or a honeycomb lattice. Here
the internal O(3) spin symmetry of the isotropic Heisenberg model is
spontaneously broken by the ground state which displays a non-zero staggered
magnetization. This system has been widely studied in condensed matter
physics and it will be instructive to compare our results with the findings
derived within the microscopic Heisenberg model. As we will see, the
effective field theory approach is by far more powerful than spin-wave theory.

Apart from our specific application concerning the partition
function of the O(3) antiferromagnet in $d$=2+1 up to three-loop
accuracy, we would like to point out that the effective field theory approach
to this condensed matter system has also proven to be very efficient in other
applications: In particular, in a recent publication on the constraint
effective potential of the staggered magnetization of the O(3) antiferromagnet
\cite{Uwe CEP}, the quantitative correctness of the magnon effective field
theory has been demonstrated in great detail at permille level accuracy by
comparison with Monte Carlo simulations of the quantum Heisenberg model using
the very efficient loop-cluster algorithm. At this accuracy, three-loop
effects clearly start manifesting themselves as there are small discrepancies
between the Monte Carlo data and the two-loop predictions of the effective
field theory. Indeed, it would be interesting to extend the finite-volume
effective field theory formulas for the constraint effective potential to
three loops and to confirm the correctness of the effective field theory
approach on an even higher level of accuracy. While this finite-volume
calculation may be performed in a future study, in the present work we focus
on finite temperature.

The paper is organized as follows. For the sake of selfconsistency, in
section \ref{method}, we give a brief outline of the effective Lagrangian
method at finite temperature. In section \ref{Feynman} we present the
evaluation of the partition function up to three-loop order in the
perturbative expansion. The issue of renormalization is then discussed in
section \ref{Divergences}. Section \ref{Results} contains our main results,
i.e., the low-temperature series for the free energy density and other
thermodynamic quantities up to three-loop order. In section
\ref{Justification} we justify why it is legitimate to use a
(pseudo-)Lorentz-invariant framework in our calculation. We then compare in
section \ref{Comparison} our results which apply to O(3) antiferromagnets in
three dimensions with those for O(3) antiferromagnets in four dimensions.
Finally, section \ref{Summary} contains our conclusions, while some technical
details concerning the renormalization and the evaluation of a specific
three-loop graph are relegated to three appendices.

\section{The effective Lagrangian method at finite temperature}
\label{method}

In a Lorentz-invariant framework the construction of effective Lagrangians is
straightforward \cite{Leutwyler foundations}: One writes down the most general
expression consistent with Lorentz symmetry and the internal, spontaneously
broken symmetry G of the underlying model in terms of Goldstone fields
$U^a(x), a = 1, \dots$, dim(G)-dim(H) -- the effective Lagrangian then
consists of a string of terms involving an increasing number of derivatives
or, equivalently, amounts to an expansion in powers of the momentum.
Furthermore, the effective Lagrangian method allows to systematically take
into account interactions which explicitly break the symmetry G of the
underlying model, provided that they can be treated as perturbations.

In the particular case we are considering, the symmetry G = O($N$) is explicitly
broken by an external field. It is convenient to collect the ($N$-1) Goldstone
fields $U^a$ in a $N$-dimensional vector $U^i = (U^0,U^a)$ of unit length,
\begin{equation}
U^i(x) \, U^i(x) \, = \, 1 \, ,
\end{equation}
and to take the constant external field along the zeroth axis,
$H^i = (H,0, \dots , 0)$. The Euclidean form of the effective Lagrangian up to
and including
order $p^4$ then reads \cite{Hasenfratz Leutwyler}:
\begin{eqnarray}
\label{Leff}
{\cal L}_{eff} & = & {\cal L}^2_{eff} + {\cal L}^4_{eff} \, , \nonumber \\
{\cal L}^2_{eff} & = &  \mbox{$ \frac{1}{2}$} F^2 {\partial}_{\mu}
U^i{\partial}_{\mu} U^i \, - \, {\Sigma}_s H^i U^i \, , \nonumber\\
{\cal L}^4_{eff} & = & \, - \, e_1 ({\partial}_{\mu} U^i
{\partial}_{\mu} U^i)^2 \, - \, e_2 \, ({\partial}_{\mu} U^i {\partial}_{\nu}
U^i)^2 + \, k_1 \! \, \frac{{\Sigma}_s}{F^2} \, (H^i U^i)
({\partial}_{\mu} U^k {\partial}_{\mu} U^k) \nonumber \\
&  & \, - k_2 \, \! \frac{{\Sigma}_s^2}{F^4} \, (H^i U^i)^2 \, - \, k_3 \,
\! \frac{{\Sigma}_s^2}{F^4} \, H^i \! H^i \, .
\end{eqnarray}
In the momentum power counting scheme, the field $U(x)$ counts as a quantity
of order 1. Derivatives correspond to one power of the momentum,
${\partial}_{\mu} \propto p$, whereas the external field $H$ counts as a term
of order $p^2$. Hence, at leading order ($\propto p^2$) we have two coupling
constants, $F$ and ${\Sigma}_s$, while at next-to-leading order
($\propto p^4$) already five constants, $e_1, e_2, k_1, k_2$ and $k_3$, show
up. Note that these couplings are not fixed by symmetry -- they parametrize
the physics of the underlying theory and have to be determined either
experimentally or in a numerical simulation. Using magnetic terminology, the
square of the effective coupling constant $F$ is the spin stiffness, while
for the O(3) antiferromagnet the quantities ${\Sigma}_s$ and $H^i$ represent
the staggered magnetization and the staggered external field, respectively.

The effective Lagrangian method provides us with a simultaneous expansion of
physical quantities in powers of the momenta and of the external field. The
essential point is that, to a given order in the low-energy expansion, only a
finite number of coupling constants and only a finite number of graphs
contribute. The leading terms stem from tree graphs, whereas loop graphs only
manifest themselves at higher orders in the derivative expansion
\cite{Weinberg}.

A crucial difference with respect to the effective analysis in four space-time
dimensions concerns the suppression of loops in Feynman graphs: While loops
are suppressed by {\it two} momentum powers in four space-time dimensions, in
three space-time dimensions loop corrections are suppressed by only {\it one}
power of momentum \cite{footnote2}. As a consequence, the number of Feynman
graphs that contribute to the perturbative expansion of the partition function
up to a given order $p^n$, will depend on the space-time dimension. As we will
see in the next section, there are fewer graphs in three dimensions that
contribute up to three-loop order.

The effective Lagrangian technique can readily be extended to finite
temperature \cite{footnote3}. In the partition function, contributions of
massive particles are suppressed exponentially, such that the Goldstone bosons
dominate the properties of the system at low temperatures. In the power
counting rules, the role of the external momenta is taken over by the
temperature, which is treated as a small quantity of order $p$. The
interaction among the Goldstone degrees of freedom in three dimensions
generates corrections of order $p/F \propto T/F$, while in four dimensions,
the corrections are of order $p^2/F^2 \propto T^2/F^2$.

In the effective Lagrangian framework at finite temperature, the partition
function is represented as a Euclidean functional integral \cite{footnote3},
\begin{equation}
\label{TempExp}
\mbox{Tr} \, [\exp(-{\cal H}/T)] \, = \, \int [{\mbox{d}}U] \,
\exp \Big( - {\int}_{\!\!\! {\cal T}} \! {\mbox{d}}^4x \, {\cal L}_{eff} \Big)
\, .
\end{equation}
The integration is performed over all field configurations which are periodic
in the Euclidean time direction, $U({\vec x}, x_4 + \beta) = U({\vec x}, x_4)$,
with $\beta \equiv 1/T$. The low-temperature expansion of the partition
function is obtained by considering the fluctuations of the field $U$ around
the ground state $V = (1, 0, \dots, 0)$, i.e. by expanding $U^0$ in powers of
$U^a$, $U^0 = \sqrt{1-U^aU^a}$. The leading contribution (order $p^2$)
contains a term quadratic in $U^a$ which describes free (pseudo-)Goldstone
bosons of mass
\begin{equation}
\label{GBMass}
M^2 = {\Sigma}_s H / F^2 \, .
\end{equation}
The remainder of the effective Lagrangian is treated as a perturbation.
Evaluating the Gaussian integrals in the standard manner, one arrives at a set
of Feynman rules which differ from the conventional rules of the effective
Lagrangian method only in one respect: the periodicity condition imposed on
the Goldstone field modifies the propagator. At finite temperature, the
propagator is given by
\begin{equation}
\label{ThermalPropagator}
G(x) \, = \, \sum_{n \,= \, - \infty}^{\infty} \Delta({\vec x}, x_4 + n \beta)
\, ,
\end{equation}
where $\Delta(x)$ is the Euclidean propagator at zero temperature. We restrict
ourselves to the infinite volume limit and evaluate the free energy density
$z$, defined by
\begin{equation}
\label{freeEnergyDensity}
z \ = \ - \, T \, \lim_{L\to\infty} L^{-3} \, \ln \, [\mbox{Tr}
\exp(-{\cal H}/T)] \, .
\end{equation}

To evaluate the graphs of the effective theory, it is convenient to use
dimensional regularization, since the symmetries of the theory are preserved
within this scheme. The zero-temperature propagator then reads
\begin{equation}
\label{regprop}
\Delta (x) \, = \, (2 \pi)^{-d} \! \int \! \! {\mbox{d}}^d p \, e^{ipx} (M^2
\! + p^2)^{-1} \, = \, {\int}_{\!\!\! 0}^{\infty} \mbox{d} \rho \, (4 \pi
\rho)^{-d/2} \, e^{- \rho M^2 - \, x^2/{4 \rho}} \, .
\end{equation}

\section{Feynman graphs}
\label{Feynman}

\begin{figure}[t]
\begin{center}
\epsfig{file=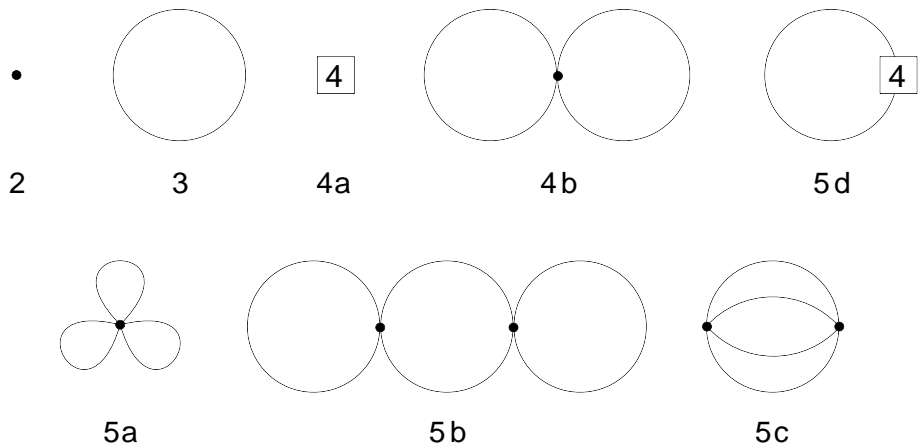,width=12cm}
\end{center}
{\caption{Feynman graphs related to the low-temperature expansion of the
partition function for an O($N$) antiferromagnet up to three-loop order in
dimension $d$=2+1. The numbers attached to the vertices refer to the piece of
the effective Lagrangian they come from. Vertices associated with the leading
term ${\cal L}^2_{eff}$ are denoted by a dot. Note that loops are suppressed
by one momentum power in $d$=2+1.}
\label{figure1} }
\end{figure}

Our aim is to evaluate the partition function of an O($N$) antiferromagnet in
dimension $d$=2+1 up to three-loop order -- the relevant Feynman graphs are
shown in Fig.\ref{figure1}. 

At leading order (order $p^2$), we have a tree graph involving
${\cal L}^2_{eff}$. The next order is $p^3$, where we have a one-loop graph.
Remember that in three dimensions every loop leads to a suppression of one
momentum power only. At order $p^4$ the next-to-leading order Lagrangian
${\cal L}^4_{eff}$ contributes to a tree graph, while the leading Lagrangian
${\cal L}^2_{eff}$ manifests itself in the form of a two-loop graph. The
situation is more involved at order $p^5$, where we have three three-loop
graphs with insertions from ${\cal L}^2_{eff}$, as well as a one-loop graph
involving ${\cal L}^4_{eff}$. Note that higher-order pieces of the effective
Lagrangian, starting with ${\cal L}^6_{eff}$, are not relevant for the
evaluation of the partition function in $d$=2+1 at the three-loop level.

\begin{figure}[t]
\begin{center}
\epsfig{file=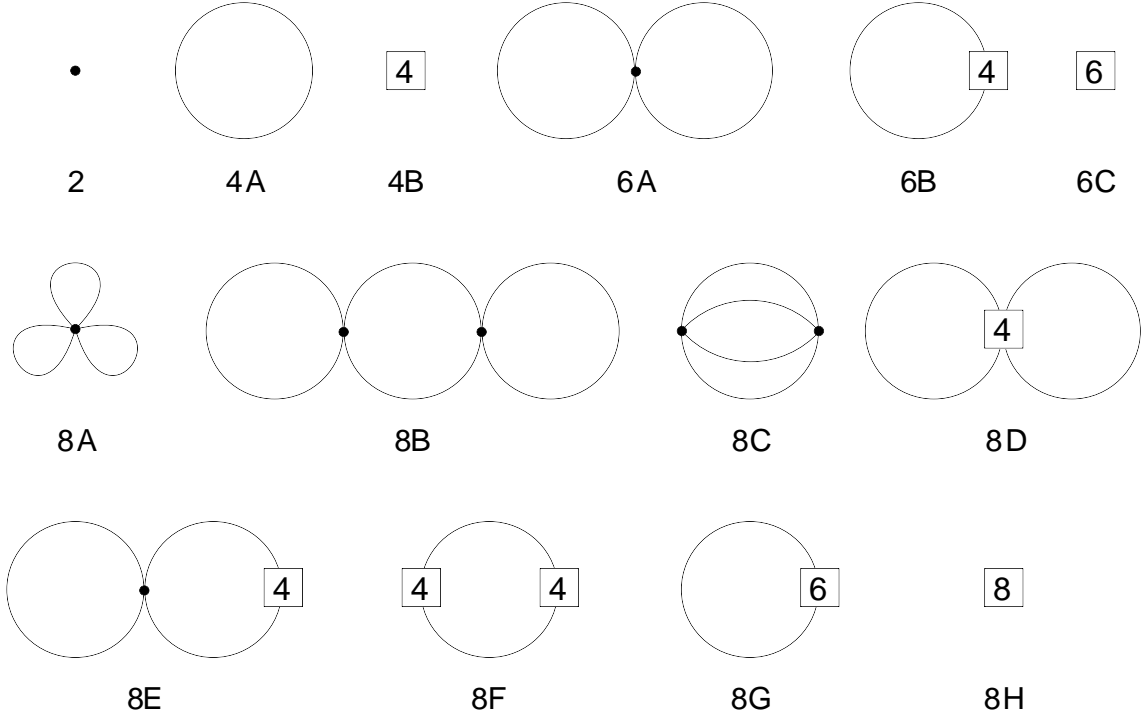,width=15cm}
\end{center}
{\caption{Feynman graphs related to the low-temperature expansion of the
partition function for an O($N$) antiferromagnet up to three-loop order in
dimension $d$=3+1. The numbers attached to the vertices refer to the piece of
the effective Lagrangian they come from. Vertices associated with the leading
term ${\cal L}^2_{eff}$ are denoted by a dot. Note that loops are suppressed
by two momentum powers in $d$=3+1. Note also that we have used capital letters
A, B, ..., H in the definition of the diagrams, in order to distinguish them
from the diagrams in $d$=2+1 displayed in Fig.\ref{figure1}, where we have
used lower-case letters.}
\label{figure2} }
\end{figure}

In order to compare our three-loop calculation in 2+1 dimensions with the one
referring to the evaluation of the partition function for an O($N$)
antiferromagnet in 3+1 dimensions, we have displayed the relevant graphs for
the latter case in Fig.\ref{figure2}. Note that loops are now suppressed by
two momentum powers, which leads to fourteen diagrams up to the three-loop
level, whereas in three dimensions we only have eight diagrams. As we will
see later on, the different loop counting and thus different organization of
Feynman graphs reflects itself also in the renormalization of these graphs
which turns out to be less involved in three dimensions.

Let us now consider the case $d$=2+1 and address the contributions from the
relevant Feynman graphs of Fig.\ref{figure1} individually. At order $p^2$, the
tree graph involving the leading-order effective Lagrangian leads to a
temperature-independent contribution,
\begin{equation}
z_2 = - F^2 M^2 \, .
\end{equation}

At order $p^3$, the one-loop graph involving ${\cal L}^2_{eff}$, which
represents the free Bose gas term, is given by
\begin{equation}
z_3 = - \mbox{$ \frac{1}{2}$} (N-1) {(4 \pi)}^{-3/2} \,
\Gamma(-\mbox{$\frac{3}{2}$}) M^3 \, - \, \mbox{$ \frac{1}{2}$}
(N-1) \, g_0(M,T) \, .
\end{equation}
The function $g_0(M,T)$ is part of a set of kinematical functions
$g_0(M,T), g_1(M,T)$ and $g_2(M,T)$ which are associated with the
$d$-dimensional noninteracting Bose gas and are defined by
\begin{equation}
\label{BoseFunctions}
g_r(M,T) \, = \, 2 {\int}_{\!\!\! 0}^{\infty}
\frac{\mbox{d} \rho}{{(4 \pi \rho)}^{d/2}} \, {\rho}^{r-1}
\, \exp(- \rho M^2) \, \sum_{n=1}^{\infty} \, \exp(-n^2 / 4 \rho T^2) \, .
\end{equation}

There are two graphs at order $p^4$. The tree graph 4a involves two coupling
constants from the next-to-leading order Lagrangian,
\begin{equation}
z_{4a} = - (k_2 + k_3) \, M^4 \, ,
\end{equation}
while the two-loop graph 4b leads to a temperature-dependent contribution
\begin{equation}
\label{graph4b}
z_{4b} = \mbox{$ \frac{1}{8}$} (N-1) (N-3) \frac{M^2}{F^2} \, {(G_1)}^2 \, .
\end{equation}
The expression $G_1$ denotes the value of the thermal propagator at the origin
\begin{equation}
G_1 \equiv G(x) |_{x=0} \, .
\end{equation}

The situation is more complicated at order $p^5$, where we have four graphs:
The three-loop graphs 5a, 5b, and 5c, involving the leading-order effective
Lagrangian ${\cal L}^2_{eff}$, as well as a one-loop graph with a vertex from
the next-to-leading order Lagrangian ${\cal L}^4_{eff}$.

Graph 5a factorizes into a term which only involves the thermal propagator at
the origin:
\begin{equation}
z_{5a} = \mbox{$ \frac{1}{16}$} (N+1) (N-1) (N-5) \frac{M^2}{F^4} \, {(G_1)}^3
\, .
\end{equation}

Likewise, graph 5b exclusively contains propagators or derivatives thereof
evaluated at the origin,
\begin{equation}
z_{5b} = - \mbox{$ \frac{1}{4}$} (N-1) {(N-3)} \frac{M^2}{F^4} \, {(G_1)}^3
\, - \, \mbox{$ \frac{1}{16}$} (N-1) {(N-3)}^2 \frac{M^4}{F^4} \, {(G_1)}^2 \,
G_2 \, .
\end{equation}
The quantity $G_2$ corresponds to an integral over the torus
${\cal T} = {\cal R}^{d_s} \times S^1$, with circle $S^1$ defined by
$- \beta / 2 \leq x_4 \leq \beta / 2$, and reads
\begin{equation}
\label{Torus}
G_2 \, = \, {\int}_{\!\!\! {\cal T}} {\mbox{d}}^d x \, \Big\{ G(x) \Big\}^2 \, .
\end{equation}
This integral can be expressed in terms of the derivative of the propagator at
the origin with respect to the mass,
\begin{equation}
G_2 \, = \, - \frac{\mbox{d} G_1}{\mbox{d} M^2} \, .
\end{equation}

Graph 5c leads to integrals over products of four propagators. Integrating by
parts, they can be brought to the form
\begin{equation}
z_{5c} = \mbox{$ \frac{1}{48}$} (N-1) (N-3) \frac{M^4}{F^4} \, J_1
\, - \, \mbox{$ \frac{1}{4}$} (N-1) (N-2) \frac{1}{F^4} \, J_2
\, + \, \mbox{$ \frac{1}{6}$} N (N-1) \frac{M^2}{F^4} \, {(G_1)}^3 \, ,
\end{equation}
where the functions $J_1$ and $J_2$ are given by
\begin{eqnarray}
J_1 & = & {\int}_{\!\!\! {\cal T}} {\mbox{d}}^d x \, \Big\{ G(x) \Big\}^4 \, ,
\nonumber \\
J_2 & = & {\int}_{\!\!\! {\cal T}} {\mbox{d}}^d x \,
\Big\{ \partial_{\mu} G(x) \, \partial_{\mu} G(x) \Big\}^2 \, .
\end{eqnarray}

Finally, the one-loop graph 5d with an insertion from ${\cal L}^4_{eff}$ yields
\begin{equation}
z_{5d} = (N-1) (k_2 - k_1) \frac{M^4}{F^2} \, G_1 \, .
\end{equation}

Collecting all the pieces, we obtain the following expression for the free
energy density of an O($N$) antiferromagnet in dimension $d$=2+1 up to and
including three loops:
\begin{eqnarray}
\label{freeEnergyDensityUnrenormalized}
z & = & - F^2 M^2
\, - \, \mbox{$ \frac{1}{2}$} (N-1) {(4 \pi)}^{-3/2} \,
\Gamma(-\mbox{$\frac{3}{2}$}) \, M^3 - \mbox{$ \frac{1}{2}$} (N-1)
\, g_0(M,T)
\, - \, (k_2 + k_3) \, M^4 \nonumber \\
& + & \mbox{$ \frac{1}{8}$} (N-1) (N-3) \frac{M^2}{F^2} \, {(G_1)}^2
\, + \, \mbox{$ \frac{1}{48}$} (N-1) (N-3) (3N-7) \frac{M^2}{F^4} \, {(G_1)}^3
\nonumber \\
& - & \mbox{$ \frac{1}{16}$} (N-1) {(N-3)}^2 \frac{M^4}{F^4} \, {(G_1)}^2 \,
G_2 \, + \, \mbox{$ \frac{1}{48}$} (N-1) (N-3) \frac{M^4}{F^4} \, J_1
\nonumber \\
& - & \mbox{$ \frac{1}{4}$} (N-1) (N-2) \frac{1}{F^4} \, J_2 + (N-1)
(k_2 - k_1) \frac{M^4}{F^2} \, G_1 + {\cal O}(p^6) \, .
\end{eqnarray}
Note that the quantities above involve the bare mass $M$ of the
(pseudo-)Goldstone bosons given in eq.(\ref{GBMass}). The thermodynamics of the
antiferromagnet is contained in the functions $g_0, G_1, G_2, J_1$, and $J_2$
which depend in a non-trivial manner on the ratio $M/T$. In the following
section, we will take care of the singularities contained in the above
expression and derive the low-temperature expansion for the free energy
density.

\section{Divergences at d=2+1 and renormalization}
\label{Divergences}

In order to analyze the divergences in the limit $d\to3$, we split the thermal
propagator into two pieces,
\begin{equation}
\label{decompositionX}
G(x) = \Delta (x) + {\bar G}(x) \, ,
\end{equation}
where $\Delta (x)$ represents the propagator at zero temperature. At the
origin, we have
\begin{eqnarray}
\label{decomposition}
G_1 & = & 2 M^2 \lambda + g_1(M,T) \, , \nonumber \\
G_2 & = & (2-d) \lambda + g_2(M,T) \, .
\end{eqnarray}
The temperature-dependent quantities $g_r(M,T)$, defined in
eq.(\ref{BoseFunctions}), are smooth functions in the limit $d\to3$. The
temperature-independent contributions involve the parameter $\lambda$
\begin{equation}
\label{lambda}
\lambda = \mbox{$\frac{1}{2}$} {(4 \pi)}^{-d/2} \,
\Gamma(1-\mbox{$\frac{d}{2}$}) \, M^{d-4} \, .
\end{equation}
Remarkably, $\lambda$ is finite in the limit $d\to3$,
\begin{equation}
\lambda = - \frac{1}{8 \pi M} \, .
\end{equation}
On the other hand, in the limit $d\to4$ the parameter $\lambda$ contains a
pole due to the singular behavior of the Gamma function. Accordingly,
logarithmic divergences in the ultraviolet show up in four space-time
dimensions, such that the next-to-leading order effective constants will
undergo a logarithmic renormalization in $d$=3+1. For the moment, however, we
focus on the case $d$=2+1.

In order to remove the singularities in the remaining integrals $J_1$ and
$J_2$, as we show in appendix A, it suffices to subtract counterterms of the
form $c_1 + c_2 \, g_1(M,T)$ and $c_3 + c_4 \, g_1(M,T)$, respectively,
\begin{eqnarray}
\label{decompositionJ}
{\bar J}_1 & = & J_1 - c_1 - c_2 \, g_1(M,T) \, , \nonumber \\
{\bar J}_2 & = & J_2 - c_3 - c_4 \, g_1(M,T) \, , 
\end{eqnarray}
where the constants $c_i$ are singular functions of the dimension $d$. While
the quantities $c_1$ and $c_3$ renormalize the vacuum energy, $c_2$ and $c_4$
renormalize the mass $M$ (see below).

We now insert the decompositions (\ref{decomposition}) and
(\ref{decompositionJ}) into the free energy density
(\ref{freeEnergyDensityUnrenormalized}) and discuss the various pieces
therein. All contributions which are independent of the temperature,
\begin{eqnarray}
\label{vacuumEnergyDensity}
z_0 & = & - F^2 M^2 - \mbox{$\frac{1}{12 \pi}$} \, (N-1) M^3 - (k_2 + k_3) \,
M^4 + \mbox{$\frac{1}{128 {\pi}^2}$} (N-1) (N-3) \frac{M^4}{F^2} \nonumber \\
& - &  \mbox{$\frac{1}{6144 {\pi}^3}$}(N-1) (N-3)(9N-23) \frac{M^5}{F^4}
 + \mbox{$ \frac{1}{48}$} (N-1) (N-3) \frac{M^4}{F^4} \, c_1 \nonumber \\
& - & \mbox{$ \frac{1}{4}$} (N-1) (N-2) \frac{1}{F^4} \, c_3
 - \mbox{$\frac{1}{4 \pi}$} \, (N-1) (k_2 - k_1) \frac{M^5}{F^2} +
{\cal O}(p^6) \, ,
\end{eqnarray}
merely renormalize the vacuum energy.

Next, we consider all terms in the free energy density
(\ref{freeEnergyDensityUnrenormalized}) which are linear in the kinematical
functions $g_r(M,T)$.  In appendix B we show that these contributions can be
merged into a single term proportional to $g_0(M_{\pi},T)$ by renormalizing the
mass, $M \to M_{\pi}$, according to
\begin{equation}
\label{massrend3}
M^2_{\pi} = M^2 + (N-3) \, \lambda \, \frac{M^4}{F^2} + \Big\{ 2 (k_2 - k_1) +
\frac{b_1}{F^2} + \frac{b_2 \, \lambda^2 M^2}{F^2} \Big\} \, \frac{M^4}{F^2}
 + {\cal O}(M^5) \, .
\end{equation}
The quantity $b_1$ is related to the singularities contained in the
coefficients $c_2$ and $c_4$. We thus see that the divergences in $b_1$,
originating from the three-loop graph 5c, are absorbed into the combination
$k_2-k_1$ (stemming from the one-loop graph 5d) of next-to-leading order
effective constants. After mass renormalization, the only surviving term
linear in the kinematical function is the contribution from the free energy
density of noninteracting magnons given by
\begin{equation}
- \mbox{$\frac{1}{2}$} \, (N-1) \, g_0(M_{\pi},T) \, ,
\end{equation}
which now depends on the renormalized mass $M_{\pi}$.

Finally we have to take care of the terms quadratic and cubic in the functions
$g_r(M,T)$, which is also done in appendix B. We are then left with the
following expression for the free energy density of an O($N$) antiferromagnet
in three dimensions,
\begin{eqnarray}
\label{freeEnergyDensityg}
z & = & z_0 - \mbox{$\frac{1}{2}$} (N-1) \, g_0
\, + \, \mbox{$\frac{1}{8}$} (N-1) (N-3) \, \frac{M_{\pi}^2}{F^2} \, {(g_1)}^2
\nonumber \\
& - & \mbox{$\frac{1}{128 \pi}$} (N-1) (N-3) (5N-11) \, \frac{M_{\pi}^3}{F^4}
\, {(g_1)}^2
\, + \, \mbox{$\frac{1}{48}$} (N-1) (N-3) (3N-7) \, \frac{M_{\pi}^2}{F^4} \,
{(g_1)}^3 \nonumber \\
& - & \mbox{$\frac{1}{16}$} (N-1) {(N-3)}^2 \, \frac{M_{\pi}^4}{F^4} \,
{(g_1)}^2 \, g_2 \, + \, \frac{Q}{F^4} + {\cal O}(p^6) \, ,
\end{eqnarray}
where we have defined the function $Q(M_{\pi},T)$ by
\begin{equation}
\label{definitionI}
Q \equiv \mbox{$ \frac{1}{48}$} (N-1) (N-3) \, M_{\pi}^4 \, {\bar J}_1
- \mbox{$ \frac{1}{4}$} (N-1) (N-2) \, {\bar J}_2 \, .
\end{equation}
The expression (\ref{freeEnergyDensityg}) for the free energy density is free
of divergences and only involves the physical mass $M_{\pi}$. In particular,
the kinematical functions are defined as $g_r = g_r(M_{\pi},T)$.

For dimensional reasons, the thermodynamic functions in
eq.(\ref{freeEnergyDensityg}) are of the form $T^p f(\tau)$, where $\tau$ is
the dimensionless ratio
\begin{equation}
\tau = \frac{T}{M_{\pi}} \, .
\end{equation}
Explicitly, in $d$=2+1 they are given by
\begin{equation}
\label{ThermalDimensionless}
g_0 = T^3 \, h_0(\tau) \, , \qquad g_1 = T \, h_1(\tau) \, , \qquad g_2 =
\frac{1}{T} \, h_2(\tau) \, , \qquad Q = T^5 \, q(\tau) \, ,
\end{equation}
such that the free energy density can be written as
\begin{eqnarray}
\label{fedTau}
z & = & z_0 - \mbox{$\frac{1}{2}$} (N-1) h_0(\tau) \, T^3
\, + \, \mbox{$\frac{1}{8}$} (N-1) (N-3) \frac{1}{F^2 {\tau}^2} {h_1(\tau)}^2
\, T^4 \nonumber \\
& - & \mbox{$\frac{1}{128 \pi}$} (N-1) (N-3) (5N-11) \frac{1}{F^4 {\tau}^3}
{h_1(\tau)}^2 \, T^5 \nonumber \\
& + & \mbox{$\frac{1}{48}$} (N-1) (N-3) (3N-7) \frac{1}{F^4 {\tau}^2}
{h_1(\tau)}^3 \, T^5 \nonumber \\
& - & \mbox{$\frac{1}{16}$} (N-1) {(N-3)}^2 \frac{1}{F^4 {\tau}^4}
{h_1(\tau)}^2 h_2(\tau) \, T^5
\, + \, \frac{1}{F^4} \, q(\tau) \, T^5 + {\cal O}(T^6) \, .
\end{eqnarray}
This expression for the free energy density of an O($N$) antiferromagnet in
2+1 dimensions represents the basic result of our paper. The ratio
$\tau=T/M_{\pi}$ can take any value, as long as the quantities $T$ and
$M_{\pi}$ themselves are small compared to the intrinsic scale $\Lambda$ of
the theory which, in the case of the O(3) antiferromagnet, may be identified
with the exchange integral $J$ of the Heisenberg model (\ref{HeisenbergModel}).

Remarkably, for $N=3$ -- the quantum Heisenberg antiferromagnet on a
square lattice -- most of the terms drop out and we are left with the
following simple expression for the free energy density of the O(3)
antiferromagnet in $d$=2+1:
\begin{equation}
\label{freeEnergyN3}
z \, = \, z_0 - h_0(\tau) \, T^3 + \frac{1}{F^4} \, q(\tau) \, T^5
+ {\cal O}(T^6) \qquad \qquad (N=3) \, .
\end{equation}
While the term cubic in the temperature corresponds to the free Bose gas, the
term proportional to five powers of the temperature represents the leading
contribution due to the spin-wave interaction. Note that for this special case
($N$=3), the function $q(\tau)$ defined in eq.(\ref{ThermalDimensionless})
only involves the contribution proportional to ${\bar J}_2$.

\section{Low-temperature series for the O(3) antiferromagnet in d=2+1}
\label{Results}

With the representation (\ref{freeEnergyN3}) for the free energy density of
the O(3) antiferromagnet in 2+1 dimensions, we are now able to discuss
various thermodynamic quantities for this system. We are particularly
interested in the limit $T \gg M_{\pi}$ which we implement by holding $T$ fixed
and sending $M_{\pi}$ (or, equivalently, the external field $H$) to zero.
Since we keep the fixed $T$ small compared to the intrinsic scale $\Lambda$ of
the underlying theory, we do not leave the domain of validity of the
low-temperature expansion.

Because the system is homogeneous, the pressure is given by the
temperature-dependent part of the free energy density,
\begin{equation}
\label{Pz}
P \, = \, z_0 - z \, = \, h_0(\tau) \, T^3 - \frac{1}{F^4} \, q(\tau) \, T^5
+ {\cal O}(T^6) \, .
\end{equation}

The non-trivial dependence of the quantity $P$ on the ratio
$\tau = T/M_{\pi}$ is contained in the functions $h_0(M_{\pi}, T)$ and
$q(M_{\pi}, T)$, which are defined in eqs.(\ref{BoseFunctions}) and
(\ref{ThermalDimensionless}). For the function $h_0(M_{\pi}, T)$ an analytical
expression can be provided in the limit $T \gg M_{\pi}$ (see appendix C):
\begin{eqnarray}
\label{expansionG0d3}
h_0^{d=3}(\tau) = \frac{1}{\pi} \Bigg[ \zeta(3) & - & \frac{1}{4}
\frac{M_{\pi}^2}{T^2} \, + \, \frac{1}{4} \frac{M_{\pi}^2}{T^2} \,
\ln\!\frac{M_{\pi}^2}{T^2}
\nonumber \\
& - & \frac{1}{6} \frac{M_{\pi}^3}{T^3} \, + \, \frac{1}{96}
\frac{M_{\pi}^4}{T^4} \, + \, {\cal O}\Big(\frac{M_{\pi}}{T}\Big)^6 \Bigg] \, .
\end{eqnarray}

\begin{figure}[t]
\begin{center}
\epsfig{file=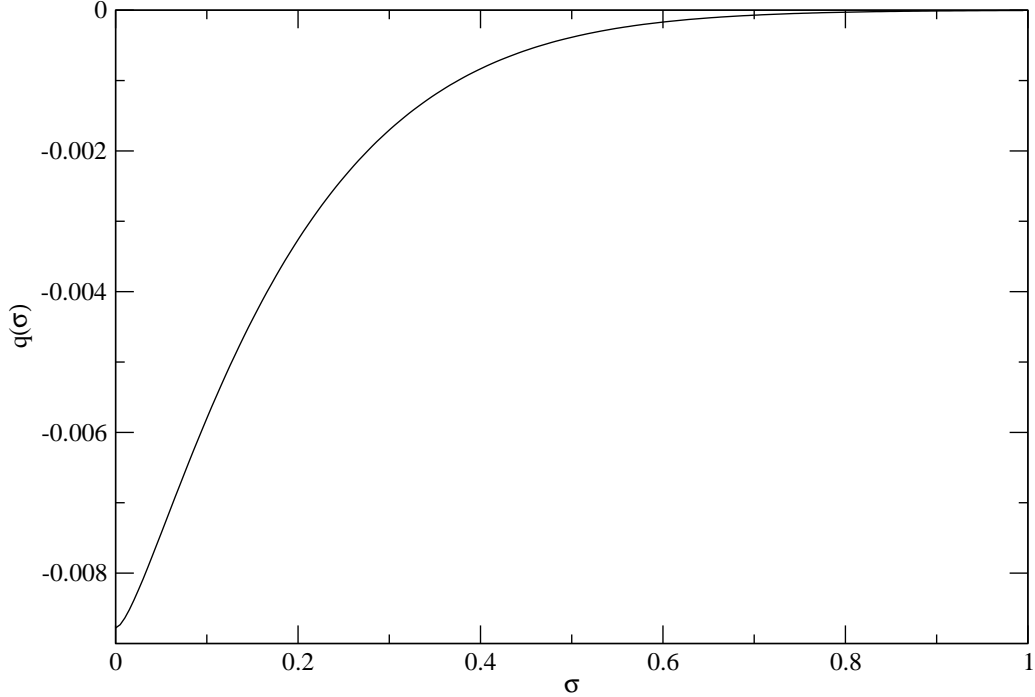,width=12cm,angle=-90}
\end{center}
{\caption{The function $q(\sigma)$ for $N$=3, where $\sigma$ is the
dimensionless parameter $\sigma=M_{\pi}/2 \pi T=1/2 \pi \tau$, introduced in
Ref.~\cite{Gerber Leutwyler}.}
\label{figure3} }
\end{figure} 

The function $q(M_{\pi}, T)$, on the other hand, we have evaluated
numerically, using the representation for ${\bar J}_2$ given in appendix A --
a plot of is provided in Fig.\ref{figure3}. Still, in the limit
$T \gg M_{\pi}$, the function may be parametrized by
\begin{equation}
q(\tau) = q_1 \; + \; q_2 \, \frac{M_{\pi}^2}{T^2} \; + \; {\cal O}
{\Big(\frac{M_{\pi}}{T}\Big)}^4 \, , \qquad \tau = \frac{T}{M_{\pi}} \, ,
\end{equation}
where the coefficients $q_i$ are real numbers. Here we only need $q_1$ which
takes the value
\begin{equation}
q_1 = -0.008779 \, . 
\end{equation}

Making use of the above representations for $h_0(M_{\pi}, T)$ and
$q(M_{\pi}, T)$, in the limit $T \gg M_{\pi}$ the pressure takes the form
\begin{equation}
\label{pressureAFd3}
P = \frac{\zeta(3)}{\pi} \, T^3 \, \Big[ 1 - \frac{\pi q_1}{\zeta(3)} \,
\frac{T^2}{F^4} + {\cal O}(T^3) \Big] \, \approx 0.3826 \, T^3 \,
\Big[ 1 + 0.02294 \, \frac{T^2}{F^4} + {\cal O}(T^3) \Big] \, . 
\end{equation}

The corresponding expressions for the energy density $u$, for the entropy
density $s$, and for the heat capacity $c_V$ for the O(3) antiferromagnet in
2+1 dimensions, are readily worked out from the thermodynamic relations
\begin{equation}
\label{Thermodynamics}
s \, = \, \frac{{\partial}P}{{\partial}T} \, , \qquad u \, = \, Ts - P \, ,
\qquad c_V \, = \, \frac{{\partial}u}{{\partial}T} \, = \, T \,
\frac{{\partial}s}{{\partial}T} \, ,
\end{equation}
with the result
\begin{eqnarray}
u & = & \frac{2 \, \zeta(3)}{\pi} \, T^3 \, \Big[ 1 -
\frac{2 \pi q_1}{\zeta(3)} \, \frac{T^2}{F^4} + {\cal O}(T^3) \Big]
\nonumber \\
& \approx & 0.7653 \, T^3 \, \Big[ 1 + 0.04589 \, \frac{T^2}{F^4}
+ {\cal O}(T^3) \Big] \, , \nonumber \\
s & = & \frac{3 \,\zeta(3)}{\pi} \, T^2 \, \Big[ 1 -
\frac{5 \pi q_1}{3 \zeta(3)} \, \frac{T^2}{F^4} + {\cal O}(T^3) \Big]
\nonumber \\
& \approx & 1.1479 \, T^2 \, \Big[ 1 + 0.03824 \, \frac{T^2}{F^4}
+ {\cal O}(T^3)\Big] \, , \nonumber \\
c_V & = & \frac{6 \, \zeta(3)}{\pi} \, T^2 \, \Big[1 -
\frac{10 \pi q_1}{3 \zeta(3)} \, \frac{T^2}{F^4} + {\cal O}(T^3) \Big] 
\nonumber \\
& \approx & 2.2958 \, T^2 \Big[ 1 + 0.07648 \, \frac{T^2}{F^4}
+ {\cal O}(T^3) \Big] \, .
\end{eqnarray}
The respective first terms in the above series represent the free Bose gas
contribution which originates from a one-loop graph. The effective interaction
among the Goldstone bosons only manifests itself through a term of order $T^5$
in the pressure, related to a three-loop graph. Interestingly, the coefficient
$q_1$ is negative, such that the magnon-magnon-interaction in the O(3)
antiferromagnet in $d$=2+1 is repulsive at low temperatures. It is remarkable
that the coefficient of the interaction term in these series is fully
determined by the symmetries inherent in the leading-order effective
Lagrangian, and does not involve any next-to-leading order coupling constants
from ${\cal L}^4_{eff}$, reflecting the anisotropies of the square lattice or
the Lorentz-noninvariant nature of the quantum Heisenberg antiferromagnet
defined on a square or a honeycomb lattice -- the symmetry is thus very
restrictive in $d$=2+1. Note that there is no interaction term of order $T^4$
in the pressure: the two-loop contribution $z_{4b}$ is proportional to $N-3$
and thus vanishes for the O(3) antiferromagnet, irrespective of the actual
value of the ratio $\tau = T/M_{\pi}$.

The fact that an interaction term proportional to four powers of the
temperature does not show up in
the temperature expansion for the pressure of the O(3) antiferromagnet in
the limit $T \gg M_{\pi}$was  already pointed out in
Ref.~\cite{Hasenfratz Niedermayer}: this was an effective Lagrangian
calculation that operated on the two-loop level. We are not aware of any
microscopic calculation that aimed at this accuracy. Moreover, our result that
the leading contribution of the magnon-magnon interaction in the pressure
is repulsive and of order $T^5$ requires a three-loop calculation on the
effective level, performed in the present study - it is probably fair to say
that this accuracy is beyond the reach of any realistic microscopic
calculation based on spin-wave theory.

In fact, as pointed out in Ref.~\cite{Hasenfratz Niedermayer}, there were
inconsistencies between the results obtained by spin-wave theory, Schwinger
boson mean field theory and Monte Carlo simulations, already with respect to
the very leading term ($\propto T^3$) in the temperature expansion of the
energy density \cite{Arovas Auerbach,Okabe et al,Barnes}. The error was later
attributed to some numerical problems in solving the equations arising in
Schwinger boson mean field theory. The systematic effective Lagrangian method,
which approaches the problem from a unified and model-independent perspective
based on the symmetries of the underlying theory, thus clearly proves to be
superior to these conventional condensed matter techniques. 

As demonstrated in a recent publication on the constraint effective potential
of the staggered magnetization of the O(3) antiferromagnet \cite{Uwe CEP},
three-loop effects clearly start manifesting themselves, as there are small
discrepancies between the very precise Monte Carlo data and the two-loop
predictions of the effective field theory. Indeed, it would be interesting to
extend the finite-volume effective field theory formulas for the constraint
effective potential to three loops, in order to confirm the correctness of the
effective field theory approach on an even higher level of accuracy and to
extract the numerical values of some combinations of effective next-to-leading
order coupling constants. Although non-trivial, this would certainly be
feasible within the framework of the systematic magnon effective field
theory.

We now make an important comment on the range of validity of the above
low-temperature series. After all, we are considering the limit
$T \gg M_{\pi}$, which we have implemented by holding $T$ fixed and sending
$M_{\pi}$, or, equivalently, the external field $H$, to zero. However,
following Mermin and Wagner \cite{Mermin Wagner}, there is no spontaneous
symmetry breaking at any finite temperature in the O(3)-invariant Heisenberg
model. Accordingly, there are no massless magnons in the low-energy spectrum
at any finite temperature. Rather, the magnons pick up an exponentially small
mass. The argument of the exponential is proportional to the inverse
temperature,
\begin{equation}
\label{npcorrelation}
m = \Big( \frac{8}{e} \Big) \, 2 \pi F^2 \,
\exp \! \Big[-\frac{2 \pi F^2}{T} \Big] \,
\Bigg\{ 1 - \frac{1}{2} \frac{T}{2 \pi F^2} + {\cal O} \Big( \frac{T^2}{F^4}
\Big) \Bigg\} \, ,
\end{equation}
as derived in
Ref.~\cite{Chakravarty Halperin Nelson,Hasenfratz Niedermayer Mass Gap}.
Strictly speaking, it is therefore not legitimate to switch off the external
field $H$ completely, because the above calculation does not take into
account the non-perturbative effect of $m$. However, the corrections due to
the non-perturbatively generated mass gap are so tiny, that they cannot
manifest themselves in the above power series. In order to verify this claim,
we now estimate the order of magnitude of these corrections.  

The above low-temperature series are valid as long as the correlation length
of the Goldstone bosons, $\xi=1/M_{\pi}$ is much smaller than the
non-perturbatively generated  correlation length $\xi_{np}=1/m$, let's say,
\begin{equation}
\label{npestimate}
\frac{1}{1000} = \frac{\xi}{\xi_{np}}
= \frac{8}{e} \, \frac{2 \pi F^2}{T} \, \frac{T}{M_{\pi}} \,
\exp \! \Big[-\frac{2 \pi F^2}{T} \Big] \, ,
\end{equation}
where we have used eq.(\ref{npcorrelation}). For the Heisenberg
antiferromagnet in $d$=2+1, the spin stiffness $F^2$ has been determined very
precisely in Monte Carlo simulations \cite{Uwe CEP,spin stiffness}.
In units of the exchange integral $J$ it takes the value
$F^2 = 0.1808(4) \, J$, such that the quantity $2 \pi F^2$ is of the order of
$J$. Now the exchange integral defines a scale in the underlying theory and
for the effective expansion to be consistent, the temperature has to be small
with respect to this scale. Assuming that
\begin{equation}
\frac{T}{2 \pi F^2} = \frac{1}{100} \, ,
\end{equation}
relation (\ref{npestimate}) then yields the ratio
\begin{equation}
\frac{M_{\pi}}{T} \approx  10^{-38} \, .
\end{equation}
Remember that, in the above low-temperature series, we have implemented the
limit $T \gg M_{\pi}$ by holding $T$ fixed and sending $M_{\pi}$ to zero. We
thus see that, in principle, we cannot completely switch off the mass
$M_{\pi}$ -- rather, we start running into trouble as soon as the ratio
$M_{\pi}/T$ is of the order of the above value. However, the error introduced
is indeed very small: the leading one-loop contribution in the free energy
density, according to eqs.(\ref{freeEnergyDensityg}) and (\ref{expansionG0d3})
is
\begin{equation}
\frac{1}{\pi}  \frac{1}{4} \frac{M_{\pi}^2}{T^2} \,
\Big[ 1 - \ln\!\frac{M_{\pi}^2}{T^2} \Big] \approx 10^{-75} \, ,
\end{equation}
which is extremely small also with respect to the three-loop contributions in
eq.(\ref{pressureAFd3}). We thus confirm that the corrections due to the
non-perturbatively generated mass gap are so tiny that they cannot manifest
themselves in the above low-temperature expansions for the thermodynamic
quantities. In other words, the subtleties raised by the Mermin-Wagner
theorem in $d$=2+1 are not relevant for our calculation.

While our effective calculation is restricted to the regime $\xi \ll \xi_{np}$,
the regime $\xi \gg \xi_{np}$, is perfectly well accessible also with
effective field theory methods. However, one has to resort to a different
type of perturbative expansion. A similar situation occurs when one considers
finite size effects: when the Goldstone boson mass is small compared to the
inverse size of the box, a different effective expansion scheme, the
so-called $\epsilon$-expansion, applies. Indeed, various problems within this
framework have been investigated in detail
\cite{epsilonExpansionGL,epsilonExpansionL,epsilonExpansionG,
epsilonExpansionMC}.
 
\section{Justification of the Lorentz-invariant framework}
\label{Justification}

In this section we would like to explain why it is justified to use a
Lorentz-invariant framework in our calculation. We have seen that anisotropies
induced by a cubic or a square lattice do not affect the leading-order
effective Lagrangian, such that ${\cal L}^2_{eff}$ can be written in a
Lorentz-invariant form, where the spin-wave velocity takes over the role of
the velocity of light: The accidental O(3) space rotation symmetry at the
${\cal L}^2_{eff}$-level implies (pseudo-)Lorentz invariance. On the other
hand, the anisotropies related to the lattice structure do show up in the
next-to-leading order effective Lagrangian ${\cal L}^4_{eff}$. Now, in our
three-loop calculation of the partition function in $d$=2+1 the only
temperature-dependent diagram involving ${\cal L}^4_{eff}$ is the one-loop
diagram 5d, which is quadratic in the magnon field. Indeed, here we have a new
term due to the lattice anisotropies, also contributing to the diagram. The
respective term,
\begin{equation}
\label{extraTerm1}
\sum_{s=1,2} \, {\partial}_s {\partial}_s U^i \, {\partial}_s {\partial}_s
U^i \, ,
\end{equation}
is invariant under the 90 degrees spatial rotation symmetry of the square
lattice, but not invariant under continuous O(3) space rotations.
Interestingly, this term is absent in the case of the honeycomb lattice, as
it is not allowed by the 60 degrees rotation symmetry.

However, both for the square and the honeycomb lattice there are additional
terms showing up at next-to-leading order: If we consider an
O(3)-symmetric, i.e., space-rotation symmetric Lagrangian ${\cal L}^4_{eff}$
-- and not a Lorentz-invariant Lagrangian ${\cal L}^4_{eff}$ as we have done
so far -- there are further terms like
\begin{equation}
\label{extraTerms2}
 \Delta U^i \Delta U^i \, , \qquad H^i U^i \partial_r U^k \partial_r U^k \, ,
\end{equation}
that also have to be taken into account in ${\cal L}^4_{eff}$. The essential
observation, however, is that all these Lorentz-noninvariant extra terms in
eqs.(\ref{extraTerm1}) and (\ref{extraTerms2}) contributing to the one-loop
graph 5d merely modify mass renormalization or give rise to higher-order
corrections of the dispersion law,
\begin{equation}
\omega(\vec k) = v |{\vec k}| + {\cal O}( {\vec k}^3)\, ,
\end{equation}
but cannot manifest themselves in the magnon-magnon interaction up to the
order $p^5$ considered in the present work. Although they give rise to an
additional term in the free energy density involving five powers of the
temperature, this is a purely kinematical effect related to the one-loop graph
5d -- the leading contribution due to the magnon-magnon interaction, also of
order $T^5$, will not be affected. Hence, our main result regarding the
weakness and the repulsive character of the magnon-magnon interaction in the
O(3) antiferromagnet in $d$=2+1 perfectly well applies to the quantum
Heisenberg antiferromagnet defined on a square or a honeycomb lattice.

\section{Antiferromagnets in 2+1 and 3+1 dimensions: Structure of the
low-temperature series}
\label{Comparison}

In this section we want to compare the low-temperature series for
antiferromagnets in 2+1 and 3+1 dimensions, pointing out differences as well
as similarities. The effective Lagrangian method is ideally suited to
understand the structure of these low-temperature series, as it adopts a
unified perspective based on symmetry considerations.

As we have discussed in Sec.\ref{Feynman}, loops in four dimensions are
suppressed by two momentum powers, whereas in three dimensions they are
suppressed by one power of momentum only. Consequently, the organization of
the loop expansion for the partition function depends on the space-time
dimension and reflects itself also in the number of Feynman diagrams that
have to be evaluated. Up to the three-loop level, we have the fourteen
diagrams in four dimensions, displayed in Fig.\ref{figure2} -- in three
dimensions there are only eight, displayed in Fig.\ref{figure1}. In
particular, in $d$=2+1, there are no two-loop graphs involving the
next-to-leading order Lagrangian ${\cal L}^4_{eff}$. Moreover, contributions
from ${\cal L}^6_{eff}$ or ${\cal L}^8_{eff}$ are not needed in $d$=2+1. One
thus notices that the restrictions imposed by symmetry are extremely strong
in $d$=2+1: Up to the three-loop level, no effective coupling constants from
${\cal L}^6_{eff}$ or ${\cal L}^8_{eff}$ enter the calculation and the
couplings in ${\cal L}^4_{eff}$ -- as we have seen -- do not affect at all
the spin-wave interaction part in the free energy density.

Another immediate consequence of the dimension-dependent loop counting is
the fact that interactions among antiferromagnetic magnons in $d$=2+1 generate
corrections of order $p/F \propto T/F$, whereas in $d$=3+1 these corrections
are of order $p^2/F^2 \propto T^2/F^2$. The low-temperature series for the
various thermodynamic quantities are thus expected to proceed in steps of one
power of $T$ in $d$=2+1 and in steps of $T^2$ in $d$=3+1 -- we will come back
to this point below.

We now briefly review the relevant results for an O($N$) antiferromagnet in
3+1 dimensions -- details of the calculation can be found in
Ref.~\cite{HofmannAF}. The formula for the pressure takes the form
\begin{equation}
\label{PressureH}
P \ = \ \mbox{$ \frac{1}{2}$} (N\!-\!1) g_0 \, + \, 4 \pi a \, (g_1)^2 \, + \,
\pi g \, \Big[ b - \frac{j}{{\pi}^3 F^4} \Big] \, + \, {\cal O} (p^{10})
\qquad \qquad (d=3+1) \, .
\end{equation}
The temperature dependence is contained in the kinematical functions
$g_r(M_{\pi}, T)$ and in $j(M_{\pi}, T)$. In the limit $H\to0$ (or,
equivalently, $T \gg M_{\pi}$) we are interested in, analytical expressions
for the functions $g_0, g_1$ and $g$ can be provided
(see Ref.~\cite{Gerber Leutwyler} or appendix C),
\begin{eqnarray}
\label{expansionGrd4}
g_0(M_{\pi},T) & = &  \mbox{$\frac{\pi^2}{45}$} \, T^4 \Bigg[ 1 \, - \,
\frac{15}{4{\pi}^2} \frac{M_{\pi}^2}{T^2}
\, + \, \frac{15}{2{\pi}^3} \frac{M_{\pi}^3}{T^3} \, + \,
\frac{45 (\gamma -\mbox{$\frac{3}{4}$}-\ln 4 \pi)}{16{\pi}^4}
\frac{M_{\pi}^4}{T^4} \, \nonumber \\
& + &  \frac{45}{32{\pi}^4} \frac{M_{\pi}^4}{T^4} \, \ln
\frac{M_{\pi}^2}{T^2} \,  + \, {\cal O}\Big(\frac{M_{\pi}}{T}\Big)^6 \Bigg]
\, , \qquad \qquad \qquad (d=3+1) \, , \nonumber \\
g_1(M_{\pi},T) & = & \mbox{$ \frac{1}{12}$} T^2 \Bigg[ 1 \, - \,
\frac{3}{\pi} \frac{M_{\pi}}{T} \, + \, {\cal O} \Big(\frac{M_{\pi}^2}{T^2}
\ln\!\frac{M_{\pi}}{T}\Big) \Bigg] \, , \nonumber\\
g(M_{\pi},T) & = & \mbox{$ \frac{1}{675}$} {\pi}^4 T^8 \Bigg[ 1 \, - \,
\frac{15}{4{\pi}^2} \frac{M_{\pi}^2}{T^2} \, + \,
{\cal O}\Big(\frac{M_{\pi}}{T}\Big)^3 \Bigg] \, ,
\end{eqnarray}
whereas the function $j$, containing the three-loop contribution from graph
8C, has to be evaluated numerically,
\begin{equation}
\label{jRepd4}
j \; = \; \nu \, \ln\!\frac{T}{M_{\pi}} \; + \; j_1 \;
+ \; j_2 \, \frac{M_{\pi}^2}{T^2} \; + \; {\cal O} \Big(\frac{M_{\pi}}{T}
\Big)^3 \, , \quad \nu \equiv \frac{5(N\!-\!1)(N\!-\!2)}{48} \, . 
\end{equation}
The coefficients $j_1$ and $j_2$ in this expansion are real numbers. Note
that the function $j(\tau)$ diverges logarithmically in the limit $H\to0$.
It should be pointed out that the renormalized or physical mass $M_{\pi}$ in
$d$=3+1 is given by 
\begin{equation}
\label{PhysicalMassD4}
M^2_{\pi} = M^2 + \, \Big\{ (N-3) \lambda  + 2 (k_2 - k_1) \Big\} \,
\frac{M^4}{F^2}  + c \frac{M^6}{F^4} + {\cal O}(M^8) \qquad \qquad (d=3+1) \, ,
\end{equation}
which is different from the analogous expression (\ref{massrend3}) in $d$=2+1,
due to the different loop counting. Without going into details, we just
mention that the quantity $\lambda$, according to eq.(\ref{lambda}), is
divergent in $d$=3+1 and that the corresponding singularity occurring in the
two-loop graph 6A (see Fig.\ref{figure2}) is absorbed into the combination
$k_2 - k_1$ of next-to-leading order coupling constants originating from
graph 6B. The absorbtion of the divergences showing up in the various graphs
of order $p^8$, on the other hand, even involves a coupling constant of
${\cal L}^6_{eff}$ stemming from graph 8G, which is contained in the quantity
$c$ of eq.(\ref{PhysicalMassD4}).

Finally, the constants $a$ and $b$ in the pressure (\ref{PressureH}) involve
the scales $H_a$ and $H_b$,
\begin{eqnarray}
\label{bConstH}
a & = & - \frac{(N\!-\!1)(N\!-\!3)}{32{\pi}} \frac{{\Sigma}_s H}{F^4}
- \, \frac{{(N\!-\!1)}^3}{256 {\pi}^3} \, \frac{{({\Sigma}_s H)}^2}{F^8}
\, \ln\!\frac{H}{H_a} \, , \nonumber\\
b & = & - \, \frac{5(N\!-\!1)(N\!-\!2)}{96 {\pi}^3 F^4} \,
\ln\!\frac{H}{H_b} \, ,
\end{eqnarray}
which are related to coupling constants of ${\cal L}^4_{eff}$ (for details see
the appendix in Ref.~\cite{HofmannAF}). The first term in $a$, linear in the
external field $H$, originates from the two-loop graph 6A. The logarithmic
contributions in $a$ and $b$ involving the two scales $H_a$ and  $H_b$,
originate from two-loop graphs with insertions from ${\cal L}^4_{eff}$. Note
that in $d$=2+1 these two-loop graphs are already beyond the
next-to-next-to-leading order considered in the present paper.

Equipped with the above formulas, the low-temperature expansion of the
pressure for an O($N$) antiferromagnet in $d$=3+1 in the limit $H\to0$
amounts to
\begin{equation}
\label{Pressure}
P \ = \ \mbox{$ \frac{1}{90}$} {\pi}^2 (N\!-\!1) \, T^4 \Bigg[ 1 \, + \,
\frac{N\!-\!2}{72} \, \frac{T^4}{F^4} \, \ln{\frac{T_p}{T}} \,
+ \, {\cal O} (T^6) \, \Bigg] \qquad \qquad (d=3+1) \, .
\end{equation}
The first contribution represents the free Bose gas term which originates from
a one-loop graph, whereas the effective interaction among the Goldstone
bosons, remarkably, only manifests itself through a term of order $T^8$. This
contribution contains a logarithm, characteristic of the effective Lagrangian
method in four space-time dimensions, which involves a scale, $T_p$, related
to $H_b$ (see the appendix in Ref.~\cite{HofmannAF}). The occurrence of a
scale involving coupling constants from ${\cal L}^4_{eff}$ is a consequence of
the space-time dimension $d$=3+1: In four dimensions the parameter $\lambda$,
defined in eq.(\ref{lambda}), contains a pole, which can be absorbed into
coupling constants of ${\cal L}^4_{eff}$ by a suitable logarithmic
renormalization. Note that the divergences in the function $j$
eq.(\ref{jRepd4}) and in the constant $b$ eq.(\ref{bConstH}) cancel, such
that the remaining expression involving the scale $T_p$ is well defined in
the limit $H \to 0$.

At low temperatures, the logarithm $\ln[T_p/T]$ in the pressure
(\ref{Pressure}) is positive, such that the interaction among the Goldstone
bosons in $d$=3+1, in the absence of an external field $H$, is repulsive,
much like in $d$=2+1. The symmetries in $d$=3+1, however, are somewhat less
restrictive than in $d$=2+1, where the interaction term -- the last term in
eq.(\ref{fedTau}) involving the function $q(\tau)$ -- is unambiguously
determined by the coupling constant $F$ of the leading order effective
Lagrangian: In $d$=3+1, next-to-leading order effective constants from
${\cal L}^4_{eff}$ do show up in the scale $T_p$. Still, the symmetry is also
rather restrictive in $d$=3+1, as it unambiguously fixes the coefficient in
front of the logarithm in terms of the coupling constant $F$. Note that there
is no term of order $T^6$ in the above series for the pressure. This is due
to the fact that the respective two-loop contribution (graph 6A,
Fig.\ref{figure2}.) in eq.(\ref{PressureH}) is proportional to the constant
$a$ that vanishes for a zero external field.

Finally, the energy density $u$, the entropy density $s$, and the heat
capacity $c_V$ in the limit $H\to0$ are given by
\begin{eqnarray}
\label{ThermodynQuantities}
u & = & \mbox{$ \frac{1}{30}$} {\pi}^2 (N\!-\!1) \, T^4 \Bigg[ 1 \, + \,
\frac{N\!-\!2}{216} \, \frac{T^4}{F^4} \Big(7 \,
\ln{\frac{T_p}{T}} - 1 \Big) \, + \, {\cal O} (T^6) \, \Bigg]
\, , \nonumber \\
s & = & \mbox{$ \frac{2}{45}$} {\pi}^2 (N\!-\!1) \, T^3 \Bigg[ 1 \, + \,
\frac{N\!-\!2}{288} \, \frac{T^4}{F^4} \Big(8 \, \ln \frac{T_p}{T}
- 1\Big) \, + \, {\cal O} (T^6) \, \Bigg] \, , \qquad \quad (d=3+1) \, ,
\nonumber \\
c_V & = & \mbox{$ \frac{2}{15}$} {\pi}^2 (N\!-\!1) \, T^3 \Bigg[ 1 \, +
\, \frac{N\!-\!2}{864} \, \frac{T^4}{F^4} \Big(56 \,
\ln{\frac{T_p}{T}} - 15 \Big) \, + \, {\cal O} (T^6) \, \Bigg] \, .
\end{eqnarray} 
Note that the limit $H \to 0$ can readily be taken in $d$=3+1, since the
Mermin-Wagner theorem does not apply here: there are no
exponentially small non-perturbative corrections in the above low-temperature
series. On general power counting grounds one would expect the
low-temperature series to proceed in steps of $T^2$ in 3+1 dimensions.
However, as for the pressure before, there are no correction terms
proportional to e.g. six powers of the temperature in the internal energy
because the constant $a$ vanishes in the limit $H \to 0$.

For the specific case $N$=3, the above series for the thermodynamic quantities
take the form
\begin{eqnarray}
P & = & \mbox{$ \frac{1}{45}$} {\pi}^2 \, T^4 \Bigg[ 1 \, + \,
\frac{1}{72} \, \frac{T^4}{F^4} \, \ln{\frac{T_p}{T}} \,
+ \, {\cal O} (T^6) \, \Bigg] \, , \qquad (d=3+1, N=3) \nonumber \\
u & = & \mbox{$ \frac{1}{15}$} {\pi}^2 \, T^4 \Bigg[ 1 \, + \,
\frac{1}{216} \, \frac{T^4}{F^4} \Big(7 \,
\ln{\frac{T_p}{T}} - 1 \Big) \, + \, {\cal O} (T^6) \, \Bigg]
\, , \nonumber \\
s & = & \mbox{$ \frac{4}{45}$} {\pi}^2 \, T^3 \Bigg[ 1 \, + \,
\frac{1}{288} \, \frac{T^4}{F^4} \Big(8 \, \ln \frac{T_p}{T}
- 1\Big) \, + \, {\cal O} (T^6) \, \Bigg] \, , \nonumber \\
c_V & = & \mbox{$ \frac{4}{15}$} {\pi}^2 \, T^3 \Bigg[ 1 \, +
\, \frac{1}{864} \, \frac{T^4}{F^4} \Big(56 \,
\ln{\frac{T_p}{T}} - 15 \Big) \, + \, {\cal O} (T^6) \, \Bigg] \, .
\end{eqnarray} 
These series, valid for the O(3) antiferromagnet in $d$=3+1, we now want to
compare with the analogous series for $d$=2+1.

As we have seen, in three dimensions the parameter $\lambda$ is finite and
therefore no such scale, involving next-to-leading order coupling constants 
from ${\cal L}^4_{eff}$, arises in the low-temperature series of the
thermodynamic quantities. In the limit $H \to 0$, the low-temperature
expansions of the thermodynamic quantities for the O(3) antiferromagnet in
$d$=2+1 take the form
\begin{eqnarray}
P & = & \frac{\zeta(3)}{\pi} \, T^3 \, \Big[ 1 - \frac{\pi q_1}{\zeta(3)} \,
\frac{T^2}{F^4} + {\cal O}(T^3) \Big] \, , \qquad (d=2+1, N=3) \, ,
\nonumber \\
u & = & \frac{2 \, \zeta(3)}{\pi} \, T^3 \, \Big[ 1 -
\frac{2 \pi q_1}{\zeta(3)} \, \frac{T^2}{F^4} + {\cal O}(T^3) \Big]
\, , \nonumber \\
s & = & \frac{3 \,\zeta(3)}{\pi} \, T^2 \, \Big[ 1 -
\frac{5 \pi q_1}{3 \zeta(3)} \, \frac{T^2}{F^4} + {\cal O}(T^3) \Big]
\, , \nonumber \\
c_V & = & \frac{6 \, \zeta(3)}{\pi} \, T^2 \, \Big[1 -
\frac{10 \pi q_1}{3 \zeta(3)} \, \frac{T^2}{F^4} + {\cal O}(T^3) \Big] \, .
\end{eqnarray}
Here we would expect the low-temperature series to proceed in steps of $T$,
since every loop in $d$=2+1 leads to an additional suppression of one power
of the temperature. However, as it was the case for $d$=3+1, there are no
next-to-leading order corrections in the above series for $d$=2+1: a term
proportional to $T^4$ in the pressure is absent. Again, the corresponding
two-loop graph 4b is proportional to $M^2_{\pi}$ (see eq.(\ref{graph4b})),
such that it vanishes in the limit $H \to 0$, much like the constant $a$
in 3+1 dimensions before. In the absence of a staggered field, the
magnon-magnon interaction both in $d$=2+1 and $d$=3+1 thus becomes very weak,
as we are dealing with a next-to-next-to-leading order effect. 

For non-zero external field, the low-temperature representations of the
thermodynamic quantities retain their form, except that the coefficients now
become functions of $M_{\pi}/T$. In the region $T \gg {M_{\pi}}$ one recovers
the results of the theory for zero external field, whereas in the opposite
limit, $T \ll {M_{\pi}}$, the gas is dilute and the particles move
non-relativistically. The properties of the system are therefore very
sensitive to the value of the ratio $M_{\pi} / T$.

To illustrate this sensitivity, let us consider the pressure and discuss the
general situation for $N>2$. In the limit $H\!\to\!0$, as we have seen, a
two-loop contribution of order $p^6$ in $d$=3+1 -- or $p^4$ in $d$=2+1 --
does not occur. This is no longer the case for an approximate symmetry
($H\!\neq\!0$): remarkably, the sign of the corresponding interaction term of
order $p^6$ ($\propto H T^4$, graph 6A) in $d$=3+1 -- or the corresponding
interaction term of order $p^4$ ($\propto H T^2$, graph 4b) in $d$=2+1 --
turns out to be negative. With respect to the limit $H\!\to\!0$, the sign of
this interaction term is thus different: in the absence of an external field,
the first non-leading term (order $p^5$ in $d$=2+1, order $p^8$ in $d$=3+1)
is positive and the interaction among the Goldstone bosons thus repulsive.
We conclude that a weak external field damps this repulsion among the
Goldstone bosons, such that the effective interaction becomes even weaker.

Interestingly, the case $N$=3 is rather special: since the two-loop
contribution both in $d$=2+1 and $d$=3+1 is proportional to ($N$-3), the
above mentioned damping of the interaction does not occur. Still, the
repulsive interaction between antiferromagnetic magnons in three or four
dimensions is very weak as we are dealing with a next-to-next-to-leading order
effect.

\section{Conclusions}
\label{Summary}

Condensed matter systems exhibiting a spontaneously broken continuous symmetry
may very efficiently be analyzed with the fully systematic method of effective
Lagrangians. In the present study we have considered O($N$) antiferromagnets
in $d$=2+1 space-time dimensions which display a spontaneously broken internal
rotation symmetry O($N$) $\to$ O($N$-1) and whose leading-order effective
Lagrangian can be brought to (pseudo-)Lorentz-invariant form. The
low-temperature properties of this system are dominated by the corresponding
Goldstone bosons, which for $N$=3 may be identified with the two
antiferromagnetic magnons or spin-wave excitations.

We have extended previous results for O($N$) antiferromagnets in $d$=2+1 to
higher orders in the derivative expansion, evaluating the partition function
up to and including three-loop diagrams. Although the renormalization and the
subsequent numerical evaluation of one particular three-loop graph turns out
to be non-trivial, the calculation is perfectly feasible within the effective
field theory framework. One of our main results is that the interaction among
magnons in the O(3)-invariant Heisenberg antiferromagnet, defined on a square
or a honeycomb lattice, is very weak and repulsive at low temperatures,
manifesting itself through a term proportional to five powers of the
temperature in the pressure. Remarkably, the coefficient of this interaction
term is fully determined by the leading-order effective Lagrangian
${\cal L}^2_{eff}$ and does not involve any higher order effective constants
from ${\cal L}^4_{eff}$ -- the symmetry is thus very restrictive in $d$=2+1.
As we have argued, additional effective constants in ${\cal L}^4_{eff}$,
taking into account the Lorentz-noninvariant nature of the system, merely
affect the renormalization of the magnon mass or yield higher-order
corrections to the magnon dispersion law, but do not affect at all the
leading contribution originating from the magnon-magnon interaction in the
pressure.

The free energy density for O($N$) antiferromagnets in $d$=3+1 up to the
three-loop level, on the other hand, does involve coupling constants from
${\cal L}^4_{eff}$, which undergo logarithmic renormalization. Accordingly,
the effective expansion of thermodynamic quantities now contains a scale,
related to these coupling constants in ${\cal L}^4_{eff}$. In four dimensions,
within the (pseudo-)Lorentz-invariant framework, we thus need more
phenomenological input, i.e. the numerical values of some next-to-leading
order effective coupling constants, in order to fully specify the structure
of the magnon-magnon interaction in the low-temperature series up to the
three-loop level. Still, the symmetry is also very restrictive here, as it
unambiguously fixes the coefficients in the expansion of the free energy
density of O($N$) antiferromagnets in $d$=3+1 up to order $T^8$, where the
logarithm involving the scale enters.

The low-temperature theorems for the various thermodynamic quantities of
O($N$) antiferromagnets in three and four space-time dimensions are exact up
to and including three loops: independently of the specific underlying model,
they are valid for any system with a spontaneously broken symmetry
O($N$) $\to$ O($N$-1), provided that the system can be described in a
(pseudo-)Lorentz-invariant framework, with the velocity of light replaced by
the spin-wave velocity. In particular, there are no approximations or
idealizations involved in our main result regarding the weakness and the
repulsive character of the magnon-magnon interaction in the O(3)
antiferromagnet in $d$=2+1: Although we use a (pseudo-)Lorentz-invariant
framework, our calculation is not just some kind of 'academic' exercise, as
the lattice anisotropies, or the Lorentz-noninvariant nature of the system
in general, cannot manifest themselves in the magnon-magnon interaction up to
the three-loop order of the perturbative expansion, considered in the present
paper --  hence our calculation applies, as it stands, to the quantum
Heisenberg antiferromagnet defined on a square or a honeycomb lattice.

We would like to emphasize that the order of the calculation presented here,
as we have argued in section \ref{Results}, appears to be beyond the
reach of any realistic microscopic calculation based on spin-wave theory 
or other standard condensed matter methods methods, such as Schwinger boson
mean field theory. The fully systematic effective Lagrangian method thus
clearly proves to be more efficient than the complicated microscopic
analysis. Another virtue of the effective Lagrangian technique is that it
addresses the problem from a unified and model-independent point of view
based on symmetry -- at large wavelengths, the microscopic structure of the
system only manifests itself in the numerical values of a few coupling
constants. Therefore the effective Lagrangian method is ideally suited to
understand similarities and differences in the structure of the
low-temperature series for antiferromagnets in three and four dimensions,
based on symmetry considerations only.

\section*{Acknowledgments}
The author would like to thank H. Leutwyler and U.-J. Wiese for stimulating
discussions and for useful comments regarding the manuscript. This work was
supported by CONACYT through Grant No. 50744-F.

\begin{appendix}

\section{Evaluation of the cateye graph in $d$=2+1 }
\label{appendixA}
\renewcommand{\theequation}{A.\arabic{equation}}
\setcounter{equation}{0}

The singularities contained in the integrals $J_1$ and $J_2$, originating from
the cateye graph 5c, may be removed by subtracting suitable counterterms.
Since our main focus is the O(3) antiferromagnet, here we only discuss the
renormalization of the function $J_2$ -- the renormalization of the quantity
$J_1$, which does not contribute to the free energy density for $N$=3
according to eq.(\ref{freeEnergyDensityUnrenormalized}), will be discussed
elsewhere.

The singularities contained in $J_2$ may be removed by subtracting the
following counterterms:
\begin{equation}
\label{FunctionsJbarJ3}
{\bar J}_2 = J_2 \, - \, c_3 \, - \, c_4 \, g_1(M,T) \, .
\end{equation}
To establish this result, we use a method, developed in
\cite{Gerber Leutwyler}, which at the same time also provides us with a
representation of the renormalized integrals suitable for numerical
evaluation. We first cut out a sphere $\cal S$ around the origin of radius
$|{\cal S}| \leq \beta/2$ and decompose $J_2$ accordingly:
\begin{equation}
J_2 = {\int}_{\!\!\! {\cal S}} \! \! {\mbox{d}}^d x \, \Big\{ \partial_{\mu}
G(x) \, \partial_{\mu} G(x) \Big\}^2 + {\int}_{\!\!\! {\cal T} \setminus {\cal S}}
\! \! {\mbox{d}}^d x \, \Big\{ \partial_{\mu} G(x) \, \partial_{\mu} G(x)
\Big\}^2 \, .
\end{equation}
In the integral over the complement ${\cal T} \! \setminus \! {\cal S}$ of
the sphere, the integrand is not singular and the limit $d\to3$ can readily
be taken. In the integral over the sphere, we insert the decomposition
(\ref{decompositionX}):
\begin{eqnarray}
\label{insertion}
J_2 & = & {\int}_{\!\!\! \cal S} \! \! {\mbox{d}}^d x \, \Bigg( \Big\{
\partial_{\mu} {\bar G} \, \partial_{\mu} {\bar G} \Big\}^2
+ 4 \partial_{\mu} \bar G \, \partial_{\mu} \bar G \, \partial_{\nu}
\bar G \, \partial_{\nu} \Delta
+ 4 \partial_{\mu} {\bar G} \, \partial_{\mu}  \Delta \,
\partial_{\nu} {\bar G} \, \partial_{\nu} \Delta \nonumber \\
& + & 2 \partial_{\mu} {\bar G} \, \partial_{\mu} {\bar G} \,
\partial_{\nu} \Delta \, \partial_{\nu} \Delta
+ 4 \partial_{\mu} {\bar G} \, \partial_{\mu} \Delta \, \partial_{\nu}
\Delta \, \partial_{\nu} \Delta
+ \Big\{ \partial_{\mu} \Delta \, \partial_{\mu} \Delta \Big\}^2 \Bigg) \, .
\end{eqnarray}
In $d$=2+1 the first four terms are convergent. However, the last two terms,
involving three and four non-thermal propagators, respectively, are divergent.

In order to extract these two singularities showing up in $d$=2+1, we follow
Ref.~\cite{Gerber Leutwyler}, where the dimension was $d$=3+1. We first
disregard derivatives and consider the expression $4 {\bar G} \, \Delta^3$,
which contains three non-thermal propagators. Since $\Delta (x)$ is Euclidean
invariant, the integral
\begin{equation}
{\int}_{\!\!\! {\cal S}} \! \! {\mbox{d}}^d x \, 4 {\bar G} \, \Delta^3
\end{equation}
only involves the angular average of ${\bar G} (x)$
\begin{equation}
f(R) = {\int} {\mbox{d}}^{d-1} \Omega \, {\bar G}(x) \, , \quad R = |x| \, .
\end{equation}
The differential equation
\begin{equation}
\Box {\bar G} = M^2 {\bar G} \,
\end{equation}
implies
\begin{equation}
\Bigg( \frac{\mbox{d}^2}{\mbox{d} R^2} + \frac{d-1}{R}
\frac{\mbox{d}}{\mbox{d} R} - M^2 \Bigg) \, f \, = \, 0 \, , \quad
R < \beta \, .
\end{equation}
Since $f$ is regular at the origin, the differential equation fixes it
uniquely up to a constant. The function $g_1 \, ch(M x_4)$ obeys the same
differential equation as ${\bar G} (x)$ and coincides with it at the origin.
The angular averages of these two quantities are therefore the same, i.e.,
\begin{equation}
{\int}_{\!\!\! {\cal S}} \! \! {\mbox{d}}^d x \, {\bar G} {\Delta}^3 =
g_1 {\int}_{\!\!\! {\cal S}} \! \! {\mbox{d}}^d x \, ch(M x_4) \, {\Delta}^3
\, .
\end{equation}
We split the integral over the sphere into two pieces,
\begin{equation}
4 g_1 {\int}_{\!\!\! {\cal S}} \! \! {\mbox{d}}^d x \, ch(M x_4) \, {\Delta}^3
= 4 g_1 {\int}_{\!\!\! {\cal R}} \! \! {\mbox{d}}^d x \,  ch(M x_4) \, {\Delta}^3
- 4 g_1 {\int}_{\!\!\! {\cal R} \setminus {\cal S}} \! \! {\mbox{d}}^d x \, 
ch(M x_4) \, {\Delta}^3 \, ,
\end{equation}
where the singularity is now contained in the integral over all Euclidean
space, in the form of the counterterm
\begin{equation}
\label{definitionC2}
c_2 = 4 {\int}_{\!\!\! {\cal R}} \! \! {\mbox{d}}^d x \, ch(M x_4) \, {\Delta}^3
\, .
\end{equation}

The same line of reasoning goes through for the expression
$4 \partial_{\mu} {\bar G} \partial_{\mu} \Delta \, \partial_{\nu} \Delta
\partial_{\nu} \Delta$ in eq.(\ref{insertion}), where one ends up with the
counterterm
\begin{equation}
\label{definitionC4}
c_4 = 4 {\int}_{\!\!\! {\cal R}} \! \! {\mbox{d}}^d x \, \partial_{\mu}
ch(M x_4) \, \partial_{\mu} \Delta \, \partial_{\nu} \Delta \partial_{\nu}
\Delta\, \, .
\end{equation}
As far as the last term in eq.(\ref{insertion}), involving four non-thermal
propagators, is concerned, it suffices to subtract the temperature-independent
integral of
$\Big\{ \partial_{\mu} \Delta (x) \partial_{\mu} \Delta (x) \Big\}^2$ over all
Euclidean space,
\begin{equation}
\label{definitionC3}
c_3 = {\int}_{\!\!\! {\cal R}} \! \! {\mbox{d}}^d x \, \Big\{ \partial_{\mu}
\Delta \partial_{\mu} \Delta \Big\}^2 \, ,
\end{equation}
in order to remove the singularity. Collecting the various pieces, we thus
arrive at the following representation for the renormalized integral in
$d$=2+1:
\begin{eqnarray}
\label{J2bar}
{\bar J}_2 & = & {\int}_{\!\!\! {\cal T}} \! \! {\mbox{d}}^3 x \, T
+ {\int}_{\!\!\! {\cal T} \setminus {\cal S}} \! \! {\mbox{d}}^3 x \, U
- {\int}_{\!\!\! {\cal R} \setminus {\cal S}} \! \! {\mbox{d}}^3 x \,
{\partial}_{\mu} \Delta {\partial}_{\mu} \Delta \cdot W \, , \nonumber \\
T & = & {\Big( {\partial}_{\mu} {\bar G} \, {\partial}_{\mu} {\bar G} \Big)}^2
+ 4 {\partial}_{\mu} {\bar G} \, {\partial}_{\mu} {\bar G} \, {\partial}_{\nu}
{\bar G} \, {\partial}_{\nu} \Delta
+ 4 {\partial}_{\mu} {\bar G} \, {\partial}_{\mu}  \Delta \, {\partial}_{\nu}
{\bar G} \, {\partial}_{\nu} \Delta \nonumber \\
& & +  2 {\partial}_{\mu} {\bar G} \, {\partial}_{\mu} {\bar G} \,
{\partial}_{\nu} \Delta \, {\partial}_{\nu} \Delta \nonumber  \, , \\
U & = & 4 \partial_{\mu} {\bar G} \partial_{\mu} \Delta \, \partial_{\nu}
\Delta \partial_{\nu} \Delta
+ \partial_{\mu} \Delta \, \partial_{\mu} \Delta \, \partial_{\nu} \Delta \,
\partial_{\nu} \Delta \, ,\nonumber \\
W & = & 4 g_1 {\partial}_{\mu} ch(M x_4) \, {\partial}_{\mu} \Delta +
{\partial}_{\mu} \Delta \, {\partial}_{\mu} \Delta \, .
\end{eqnarray}
This expression involves ordinary, convergent integrals. Exploiting the fact
that ${\bar G}(x)$ and $\Delta(x)$ only depend on $r=|{\vec x}|$ and on
$t=x_4$, the integrals occurring in this representation become effectively
two-dimensional
\begin{equation}
{\mbox{d}}^3 x = 2 \pi r \, {\mbox{d}} r \, {\mbox{d}} t \, .
\end{equation}
Note that the quantity ${\bar J}_2$ must be independent of the size of the
sphere -- this provides us with a welcome numerical consistency check of our
calculation.

It is instructive to compare our decomposition of the integrals (\ref{J2bar})
with the decomposition originally used in Ref.~\cite{Gerber Leutwyler}, which
in $d$=2+1 amounts to
\begin{eqnarray}
\label{J2barGL}
{\bar J}_2 & = & {\int}_{\!\!\! {\cal T} \setminus {\cal S}} \! \! {\mbox{d}}^3 x
\, {\widetilde U}
+ {\int}_{\!\!\! {\cal S}} \! \! {\mbox{d}}^3 x \, {\widetilde V}
- {\int}_{\!\!\! {\cal R} \setminus {\cal S}} \! \! {\mbox{d}}^3 x \,
{\partial}_{\mu} \Delta {\partial}_{\mu} \Delta \cdot {\widetilde W} \, ,
\nonumber \\
{\widetilde U} & = & {\Big( {\partial}_{\mu} G \, {\partial}_{\mu} G \Big)}^2
\, , \nonumber \\
{\widetilde V} & = & {\Big( {\partial}_{\mu} {\bar G} \, {\partial}_{\mu}
{\bar G} \Big)}^2
+ 4 {\partial}_{\mu} {\bar G} {\partial}_{\mu} {\bar G} \, {\partial}_{\nu}
{\bar G} \, {\partial}_{\nu} \Delta + 2 Q_{\mu \mu } \, {\partial}_{\nu}
\Delta \, {\partial}_{\nu} \Delta + 4 Q_{\mu \nu } \, {\partial}_{\mu} \Delta
\, {\partial}_{\nu} \Delta \, ,\nonumber \\
{\widetilde W} & = & {\widetilde w} +  4 g_1 {\partial}_{\mu} ch(M x_4) \,
{\partial}_{\mu} \Delta + {\partial}_{\mu} \Delta \, {\partial}_{\mu} \Delta \, ,
\end{eqnarray}
with 
\begin{eqnarray}
{\widetilde w} & = &  \frac{1}{x^2} \Big[ ( \mbox{$\frac{3}{2}$} x^4 -
\mbox{$\frac{9}{2}$} x^2 x_4^2 + 9 x^4_4 ) \, g^2_0 + 12 M^2 x^4_4 \, g_0
\, g_1 + 2 ( 2M^4 x^4_4 + M^4 x^2 x^2_4 ) \, g^2_1 \Big] \, ,
\nonumber \\
Q_{\mu \nu} & = & {\partial}_{\mu} {\bar G}(x) \, {\partial}_{\nu} {\bar G}(x)
- {\bar G}_{\mu \alpha} {\bar G}_{\nu \beta} \, x_{\alpha} x_{\beta} \, ,
\nonumber \\
{\bar G}_{\mu \nu} & = & - \mbox{$\frac{1}{2}$} \delta_{\mu \nu} \, g_0 +
\delta^4_{\mu}  \delta^4_{\nu} \, ( \mbox{$\frac{3}{2}$} g_0 + M^2 g_1) \, .
\end{eqnarray}
The main difference between the two decompositions (\ref{J2bar}) and
(\ref{J2barGL}) concerns the terms involving two thermal propagators, where we
have
\begin{eqnarray}
\label{J2comparison}
& & {\int}_{\!\!\! {\cal S}} \! \! {\mbox{d}}^3 x \, \Big( 2 {\partial}_{\mu}
{\bar G} \, {\partial}_{\mu} {\bar G} \, {\partial}_{\nu} \Delta \,
{\partial}_{\nu} \Delta
+ 4 {\partial}_{\mu} {\bar G} \, {\partial}_{\mu}  \Delta\, {\partial}_{\nu}
{\bar G} \, {\partial}_{\nu} \Delta \Big) \nonumber \\
& & = {\int}_{\!\!\! {\cal S}} \! \! {\mbox{d}}^3 x \, \Big( 2 Q_{\mu \mu } \, 
{\partial}_{\nu} \Delta \, {\partial}_{\nu} \Delta + 4 Q_{\mu \nu } \,
{\partial}_{\mu} \Delta \, {\partial}_{\nu} \Delta \Big)
- {\int}_{\!\!\! {\cal R} \setminus {\cal S}} \! \! {\mbox{d}}^3 x \,
{\partial}_{\mu} \Delta \, {\partial}_{\mu} \Delta \,\cdot {\widetilde w}
\nonumber \\
& & + {\int}_{\!\!\! {\cal R}} \! \! {\mbox{d}}^3 x \, \Big(
2 {\bar G}_{\mu \alpha} {\bar G}_{\mu \beta}  \, x_{\alpha} x_{\beta} \,
{\partial}_{\nu} \Delta \, {\partial}_{\nu} \Delta 
+ 4 {\bar G}_{\mu \alpha} {\bar G}_{\nu \beta}  \, x_{\alpha} x_{\beta} \,
{\partial}_{\mu} \Delta \, {\partial}_{\nu} \Delta \Big) \, .
\end{eqnarray}
Now, in four dimensions the integral over the sphere on the left hand side
contains a logarithmic singularity -- this was the reason why in
Ref.~\cite{Gerber Leutwyler} the above decomposition was performed: the
singularity then occurs again on the right hand side in the integral over all
Euclidean space. It turns out that, in order to renormalize the integral
$J_2$ in $d$=3+1, it is thus not sufficient to just subtract the two
counterterms $c_3$ and $c_4$ in eq.(\ref{FunctionsJbarJ3}) -- rather, one has
to subtract two more terms,
\begin{equation}
{\bar J_2} = J_2 -c_3 -c_4 \, g_1 + \mbox{$\frac{1}{3}$} (d+6)(d-2) \, \lambda
\, {({\bar G}_{\mu \nu} )}^2 + \mbox{$\frac{2}{3}$} (d-2) \, \lambda M^4
{(g_1)}^2 \, ,
\end{equation} 
in order to remove all singularities in $J_2$.

In three dimensions, as we have seen, the integral over the sphere on the left
hand side of eq.(\ref{J2comparison}) is not singular. Likewise the integral
over all Euclidean space on the right hand side is perfectly well defined, 
such that there is no need to introduce the above decomposition
(\ref{J2barGL}) in three dimensions in the first place. Still, in order to
check that our entire calculation is consistent, we have verified both
analytically and numerically that the evaluation of the quantity ${\bar J}_2$
via eqs.(\ref{J2bar}) and (\ref{J2barGL}) yields the same result.

We close this section with a comment regarding dimensional regularization and
the different structure of the singularities in three and four dimensions,
respectively. The essential point can be seen in the identity
(\ref{J2comparison}) which involves two thermal propagators. In four
dimensions, as we have seen, the cateye graph is of order $p^8$ and so is the
singularity occurring in the last term of eq.(\ref{J2comparison}). Now, in
four dimensions we also have two-loop graphs which are of the same order
$p^8$: graphs 8D and 8E (see Fig.\ref{figure2}) which involve vertices from
the next-to-leading order effective Lagrangian ${\cal L}^4_{eff}$. Therefore,
the singularities in the last term of eq.(\ref{J2comparison}) -- the integral
over all Euclidean space -- can be absorbed into a combination of these
next-to-leading order coupling constants.

In three dimensions, on the other hand, these two-loop graphs are of order 
$p^6$, i.e., beyond the order $p^5$ considered in the present work. Loosely
speaking, there is no communication between the cateye graph (order $p^5$) and
these two-loop graphs of order $p^6$ and it so seems that -- in three
dimensions -- the 'singularities' in the last term of eq.(\ref{J2comparison})
cannot be absorbed, as there are no next-to-leading order coupling constants
available. However, in three dimensions the parameter $\lambda$ arising in
the last term of eq.(\ref{J2comparison}) is finite, such that the bookkeeping
of 'divergences' in three dimensions is perfectly consistent.

\section{Renormalization}
\label{appendixB}
\renewcommand{\theequation}{B.\arabic{equation}}
\setcounter{equation}{0}

In appendix B, we would like to derive the expression
(\ref{freeEnergyDensityg}) for the free energy density of an O($N$)
antiferromagnet in $d$=2+1. We first show that all the terms in the free
energy density (\ref{freeEnergyDensityUnrenormalized}) that are linear in the
kinematical functions $g_r(M,T)$ can be merged into a single such function,
namely $g_0$, by replacing the bare mass $M$ with the physical mass $M_{\pi}$.

With the decompositions (\ref{decomposition}) and (\ref{decompositionJ}), the
terms in the free energy density (\ref{freeEnergyDensityUnrenormalized})
linear in $g_r(M,T)$ read
\begin{eqnarray}
\label{gLinear}
z^{\{1\}} & = & - \mbox{$\frac{1}{2}$} (N-1) \, g_0(M,T) \, + \,
\mbox{$\frac{1}{2}$} (N-1) (N-3) \, \frac{M^4}{F^2} \, \lambda \, g_1(M,T)
\nonumber \\
& - &\mbox{$\frac{1}{4}$} (N-1) {(N-3)}^2 \, \frac{M^8}{F^4} \, \lambda^2 \,
g_2(M,T) + (N-1) (k_2-k_1) \, \frac{M^4}{F^2} \, g_1(M,T) \nonumber \\
& + & \mbox{$\frac{1}{48}$} c_2 (N-1) (N-3) \, \frac{M^4}{F^4} \, g_1(M,T) -
\mbox{$\frac{1}{4}$} c_4 (N-1) (N-2) \, \frac{1}{F^4} \, g_1(M,T) \nonumber \\
& + & \mbox{$\frac{1}{2}$} (N-1) (N-3)(2N-5) \, \frac{M^6}{F^4} \, \lambda^2
g_1(M,T) \, .
\end{eqnarray}
Now the pressure at low temperatures is of order $\exp(-M_{\pi}/T)$
originating from one particle states -- states containing two or more
Goldstone bosons only show up at order $\exp(-2M_{\pi}/T)$. Therefore it is
possible to extract the physical Goldstone boson mass $M_{\pi}$ from the
behavior of the pressure at low temperatures,
\begin{equation}
M_{\pi} = - \lim_{T\to0} \, T \, \ln P \, .
\end{equation}
Using the relation
\begin{equation}
g_{r+1} = - \frac{\mbox{d} g_r}{\mbox{d} M^2} \, ,
\end{equation}
this limit amounts to
\begin{equation}
\label{MassRenormalization}
M^2_{\pi} = M^2 + (N-3) \, \lambda \, \frac{M^4}{F^2} + \Big\{ 2 (k_2 - k_1) +
\frac{b_1}{F^2} + \frac{b_2 \, \lambda^2 M^2}{F^2} \Big\} \, \frac{M^4}{F^2}
 + {\cal O}(M^5) \, ,
\end{equation}
where the coefficients $b_1$ and $b_2$ are given by
\begin{eqnarray}
b_1 & = & \mbox{$\frac{1}{24}$} (N-3) \, \gamma_2 - \mbox{$\frac{1}{2}$} (N-2)
\, \gamma_4 \, , \nonumber \\
b_2 & = & (N-3)(2N-5) \, .
\end{eqnarray}
The quantities $\gamma_2$ and $\gamma_4$ are singular functions of the
dimension $d$ and are related to the coefficients $c_2$ and $c_4$ -- defined in
eqs.(\ref{definitionC2}) and (\ref{definitionC4}) -- as follows:
\begin{equation}
c_2 = \gamma_2 M^{2d-6} \, , \quad c_4 = \gamma_4 M^{2d-2} \, .
\end{equation}
Inspecting the curly bracket in formula (\ref{MassRenormalization}) one thus
notices that the infinities contained in $c_2$ and $c_4$, which stem from the
three-loop graph 5c, are absorbed into the combination $k_2-k_1$ of
next-to-leading order coupling constants, originating from the one-loop graph
5d. Note that in $d$=2+1 the parameter $\lambda$ is finite, such that second
term on the right hand side of eq.(\ref{MassRenormalization}), coming from
the two-loop graph 4b, does not contain any singularities.

One readily verifies that the replacement $g_0(M,T) \to g_0(M_{\pi},T)$ in the
first term of eq.(\ref{gLinear}) cancels all other terms linear in $g_r(M,T)$.
The free energy density, linear in the kinematical functions, thus takes the
simple form
\begin{equation}
- \mbox{$\frac{1}{2}$} (N-1) \, g_0(M_{\pi},T) \, .
\end{equation}

We now proceed with the terms in the free energy density that are quadratic in
the kinematical functions $g_r(M,T)$. They are
\begin{eqnarray}
\label{freeEnergyDensityQuadratic}
z^{\{2\}} & = & \mbox{$\frac{1}{8}$} (N-1) (N-3) \, \frac{M^2}{F^2} \,
g_1(M,T)^2 + \mbox{$\frac{1}{16}$} (N-1) (N-3) (7N-17) \, \frac{M^4}{F^4} \,
\lambda \, g_1(M,T)^2 \nonumber \\
& - & \mbox{$\frac{1}{4}$} (N-1) {(N-3)}^2 \, \frac{M^6}{F^4} \, \lambda \,
g_1(M,T) \, g_2(M,T) \, .
\end{eqnarray}
In the first term we make the replacement $g_1(M,T) \to g_1(M_{\pi},T)$, which
amounts to
\begin{equation}
g_1(M_{\pi},T)^2 = g_1(M,T)^2 - \Big\{ 2(N-3) \frac{M^4}{F^2} \, \lambda  +
{\cal O}(M^4) \Big\} \, g_1(M,T) \, g_2(M,T) \, .
\end{equation}
One notices that this cancels the third term in
eq.(\ref{freeEnergyDensityQuadratic}). We are thus left with
\begin{equation}
\mbox{$\frac{1}{8}$} (N-1) (N-3) \, \frac{M^2}{F^2} \, g_1(M_{\pi},T)^2 +
\mbox{$\frac{1}{16}$} (N-1) (N-3) (7N-17) \, \frac{M_{\pi}^4}{F^4} \, \lambda
\, g_1(M_{\pi},T)^2 \, .
\end{equation}
Note that in the second term we have also replaced the bare mass by the
physical mass, both in the prefactor and in the kinematical function: this is
legitimate as the difference is beyond our accuracy. Finally, in the
prefactor of the first term, we also express the bare mass by the physical
mass using relation (\ref{MassRenormalization}), obtaining the following
terms in the free energy density quadratic in the kinematical functions:
\begin{equation}
\mbox{$\frac{1}{8}$} (N-1) (N-3) \, \frac{M_{\pi}^2}{F^2} \, g_1(M_{\pi},T)^2
+ \mbox{$\frac{1}{16}$} (N-1) (N-3) (5N-11) \, \frac{M_{\pi}^4}{F^4} \,
\lambda \, g_1(M_{\pi},T)^2 \, .
\end{equation}

To end up, we take care of the remaining terms in the free energy density
that are either cubic in the kinematical functions $g_r(M,T)$ or are related
to integrals over the torus:
\begin{eqnarray}
z^{\{3\}} & = & \mbox{$\frac{1}{48}$} (N-1) (N-3) (3N-7) \, \frac{M^2}{F^4} \,
g_1(M,T)^3 \nonumber \\
& - & \mbox{$\frac{1}{16}$} (N-1) {(N-3)}^2 \, \frac{M^4}{F^4} \, g_1(M,T)^2
\,  g_2(M,T) \nonumber \\
& + & \mbox{$ \frac{1}{48}$} (N-1) (N-3) \, \frac{M^4}{F^4} \, {\bar J}_1
- \mbox{$ \frac{1}{4}$} (N-1) (N-2) \, \frac{1}{F^4} \, {\bar J}_2 \, .
\end{eqnarray}
Again, we replace the bare mass by the physical mass in the above terms, both
in the kinematical functions and in the prefactors, as the difference is
beyond our accuracy. No cancellations of terms occur here.

Collecting the various contributions, we arrive at the expression for the
free energy density of an O($N$) antiferromagnet in $d$=2+1:
\begin{eqnarray}
z & = & z_0 - \mbox{$\frac{1}{2}$} (N-1) \, g_0
\, + \, \mbox{$\frac{1}{8}$} (N-1) (N-3) \, \frac{M_{\pi}^2}{F^2} \,
{(g_1)}^2 \nonumber \\
& - & \mbox{$\frac{1}{128 \pi}$} (N-1) (N-3) (5N-11) \, \frac{M_{\pi}^3}{F^4}
\, {(g_1)}^2 \, + \, \mbox{$\frac{1}{48}$} (N-1) (N-3) (3N-7) \,
\frac{M_{\pi}^2}{F^4} \, {(g_1)}^3 \nonumber \\
& - & \mbox{$\frac{1}{16}$} (N-1) {(N-3)}^2 \, \frac{M_{\pi}^4}{F^4} \,
{(g_1)}^2 \, g_2
\, + \, \mbox{$ \frac{1}{48}$} (N-1) (N-3) \, \frac{M_{\pi}^4}{F^4} \,
{\bar J}_1 \nonumber \\
& - & \mbox{$ \frac{1}{4}$} (N-1) (N-2) \, \frac{1}{F^4} \, {\bar J}_2
+ {\cal O}(p^6) \, .
\end{eqnarray}
Note that only the physical mass $M_{\pi}$ occurs in the above formula: in
particular, the kinematical functions are $g_r = g_r(M_{\pi},T)$. Remember
that the temperature-independent contribution $z_0$ is the vacuum energy
density given in eq.(\ref{vacuumEnergyDensity}).

\section{Properties of the kinematical functions
$g_r(M,T)$}
\label{appendixC}
\renewcommand{\theequation}{C.\arabic{equation}}
\setcounter{equation}{0}

In this appendix we discuss some properties of the kinematical functions
$g_r(M,T)$,
defined by
\begin{equation}
\label{BoseFunctionsAppendix}
g_r(M,T) \, = \, 2 {\int}_{\!\!\! 0}^{\infty}
\frac{\mbox{d} \rho}{{(4 \pi \rho)}^{d/2}} \, {\rho}^{r-1}
\, \exp(- \rho M^2) \, \sum_{n=1}^{\infty} \, \exp(-n^2 / 4 \rho T^2) \, .
\end{equation}
We follow the appendices A and B of Ref.~\cite{Hasenfratz Leutwyler} where the
analogous kinematical functions for finite volume were considered. Here we
adapt the method described therein to finite temperature.

We are particularly interested in the expansion of $g_r(M,T)$ in the limit
$M \to 0$, where infrared singularities occur. We introduce the Jacobi
theta-function,
\begin{equation} \label{theta}
S(x) = \sum_{n=-\infty}^{\infty} \, e^{- \pi n^2 x} \, ,
\end{equation}
and express the kinematical functions (\ref{BoseFunctionsAppendix}) by
\begin{equation}
\label{JacobiRep}
g_r(M,T) = \frac{T^{d-2r}}{(4 \pi)^r} \int_{0}^{\infty} \, dt \,
t^{r-\frac{d}{2}-1} \, \exp \left(- \frac{M^2 t}{4 \pi T^2} \right) \,
\Big[ S(1/t) - 1 \Big] \ .
\end{equation}

The integration is split into two regions, $0 \ \le t \le 1$ and $1 \le t <
\infty$. In the second region we use the identity
\begin{equation}
\label{Poisson}
S(x) = \frac{1}{\sqrt{x}} \, S(1/x) \, ,
\end{equation}
change the integration variable $t \to 1/t$, and arrive at the following
representation for the kinematical functions $g_r(M,T)$:
\begin{equation}
\label{repgr}
g_r(M,T) =   \frac{T^{d-2r}}{(4 \pi)^r} \, \Big\{ {\tilde a}_r +
b_{r-\frac{d}{2}+\frac{1}{2}} - b_{r-\frac{d}{2}} \Big\}  \ ,
\end{equation}
with
\begin{equation}
{\tilde a}_r = \int_0^1 dt \, t^{r-\frac{d}{2}-1} \, \exp
\left( - \frac{M^2 t}{4 \pi T^2}  \right) \, 
\Big[ S(1/t) - 1 \Big] + \int_0^1 dt \, t^{-r+\frac{d}{2}-\frac{3}{2}} \, \exp
\left( - \frac{M^2}{4 \pi T^2 t}  \right) \, \Big[ S(1/t) - 1 \Big] 
\end{equation}
and
\begin{equation}
b_s = \int_1^{\infty} dt \, t^{s-1} \, \exp \left( - \frac{M^2 t}{4 \pi T^2}
\right)  \ .
\end{equation}

The function ${\tilde a}_r$ does not contain infrared singularities and the
expansion in powers of $M^2$ is of the form
\begin{equation}
\label{repar}
{\tilde a}_r = \sum_{n=0}^{\infty} {\Big( - \frac{M^2}{4 \pi T^2} \Big)}^n \,
\frac{1}{n!} \, 
\Big\{ {{\hat \alpha}}_{r+n-\frac{d}{2}} + {{\hat \alpha}}_{-r-n+\frac{d}{2}-
\frac{1}{2}} \Big\} \, ,
\end{equation}
where
\begin{equation}
\label{repap}
{\hat \alpha}_p = \int_{0}^{1} dt \, t^{p-1} \, \Big\{ S(1/t) - 1 \Big\} \ .
\end{equation}

The infrared singularities are contained in the incomplete $\Gamma$-function
$b_s$:
\begin{eqnarray}
b_s & = & {\Big( \frac{M^2}{4 \pi T^2} \Big)}^{-s}  \, \Gamma(s) \, - \,
\sum_{n=0}^{\infty}  \frac{1}{n!} {\Big( - \frac{M^2}{4 \pi T^2} \Big)}^n
\frac{1}{n+s} \, .
\end{eqnarray}
The pole in the function $\Gamma(s)$ at $s=0,-1,-2, \dots$ is compensated by
a pole occurring in the second piece of $b_s$ which is analytic in $M$. The
two singularities can be merged and one ends up with a logarithmic
contribution. Details can be found in Ref.~\cite{Hasenfratz Leutwyler} --
here we give the explicit expression for $b_{-N} \, (N \ge 0)$,
\begin{eqnarray}
b_{-N} & = & \frac {(-1)^{N+1}}{N!} \, \Big( \frac{M^2}{4 \pi T^2} \Big)^N \,
\Big\{ \log  \frac{M^2}{4 \pi T^2} + \gamma
- \sum_{n=1}^N \frac{1}{n} \Big\} \nonumber \\
& + & \sum_{n\neq N} \, \frac{1}{n!} \, \Big(- \frac{M^2}{4 \pi T^2} \Big)^n \,
\frac{1}{N-n} \ ,
\end{eqnarray}
where the quantity $\gamma \approx 0.577$ denotes Euler's constant. Note that,
in the second sum over $n$, the value $n=N$ is to be omitted. Moreover, for
$N$=0 the sum in the curly bracket is to be omitted.

We now consider the kinematical function $g_0(M,T)$ both in three and four
dimensions. According to eq.(\ref{repgr}) we have
\begin{eqnarray}
\label{kinematical}
g^{d=3}_0(M,T) & = & T^3 \, \Big\{ {\tilde a}_0 + b_{-1} - b_{-\frac{3}{2}}
\Big\} \, , \nonumber \\
g^{d=4}_0(M,T) & = & T^4 \, \Big\{ {\tilde a}_0 + b_{-\frac{3}{2}} - b_{-2}
\Big\} \, .
\end{eqnarray}
Using relation (\ref{Poisson}) and the identity
\begin{equation}
{\pi}^{-z/2} \, \Gamma(z/2) \, \zeta(z) = \frac{1}{2} \int_{0}^{\infty} dt \,
t^{\frac{z}{2}-1} \, \Big\{ S(t) - 1 \Big\} \, ,
\end{equation}
one readily shows that the various contributions in eq.(\ref{kinematical}) can
be merged into a single series in $M$, involving Riemann zeta functions,
\begin{eqnarray}
g_0^{d=3}(M,T) & = & T^3 \, \Bigg[ \frac{\zeta(3)}{\pi} + \frac{1}{4 \pi}
\frac{M^2}{T^2} \, \Big\{ \ln\!\frac{M^2}{T^2} -1 \Big\} - \frac{1}{6 \pi}
\frac{M^3}{T^3}
\nonumber \\
& + & 2 \pi \, \sum_{n=2}^{\infty} \frac{(-1)^n}{n!} \, \Big(  \frac{M}{2 \pi T}
\Big)^{2n} \, \Gamma(n-1) \, \zeta(2n-2) \Bigg] \, , 
\qquad \qquad \quad (T \gg M) \, , \nonumber \\
g_0^{d=4}(M,T) & = & T^4 \, \Bigg[ \frac{\pi^2}{45} \, - \, \frac{1}{12}
\frac{M^2}{T^2} \, + \, \frac{1}{6 \pi} \frac{M^3}{T^3}
\, + \,  \frac{ (2\gamma -\mbox{$\frac{3}{2}$} )}{32{\pi}^2} \frac{M^4}{T^4} \, 
+ \frac{1}{32{\pi}^2} \, \frac{M^4}{T^4} \, \ln \frac{M^2}{16 \pi^2 T^2}
\nonumber \\
& + & 2 \pi^{3/2} \, \sum_{n=3}^{\infty} \frac{(-1)^n}{n!} \,
\Big(  \frac{M}{2 \pi T} \Big)^{2n} \, \Gamma(n-\mbox{$\frac{3}{2}$}) \,
\zeta(2n-3) \Bigg]  \, .
\end{eqnarray}
Explicitly, the first few terms in these series read
\begin{eqnarray}
g_0^{d=3}(M,T) & = & \frac{1}{\pi} T^3 \Bigg[ \zeta(3) - \frac{1}{4}
\frac{M^2}{T^2} \, + \, \frac{1}{4} \frac{M^2}{T^2} \,
\ln\!\frac{M^2}{T^2}
\nonumber \\
& - & \frac{1}{6} \frac{M^3}{T^3} \, + \, \frac{1}{96}
\frac{M^4}{T^4} \, + \, {\cal O}\Big(\frac{M}{T}\Big)^6 \Bigg] \, , 
\qquad \qquad \qquad (T \gg M) \, , \nonumber \\
g_0^{d=4}(M,T) & = & \frac{\pi^2}{45} \, T^4 \Bigg[ 1 \, - \,
\frac{15}{4{\pi}^2} \frac{M^2}{T^2} \, + \, \frac{15}{2{\pi}^3} \frac{M^3}{T^3}
\, + \,  \frac{45 (\gamma -\mbox{$\frac{3}{4}$}-\ln 4 \pi)}{16{\pi}^4}
\frac{M^4}{T^4} \, \nonumber \\
& + &  \frac{45}{32{\pi}^4} \frac{M^4}{T^4} \, \ln \frac{M^2}{T^2} \,  + \,
{\cal O}\Big(\frac{M}{T}\Big)^6 \Bigg] \, .
\end{eqnarray}
The expansion for the function $g_0^{d=4}(M,T)$ coincides with the expression
derived in the appendix of Ref.~\cite{Gerber Leutwyler}, where a different
method was used. The series for the kinematical functions $g_1(M,T)$ and
$g_2(M,T)$ can readily be obtained using the relation
\begin{equation}
g_{r+1} = - \frac{\mbox{d} g_r}{\mbox{d} M^2} \, .
\end{equation}

\end{appendix}


\begin{thebibliography}{10}

\bibitem{Hasenfratz Leutwyler}
P.\ Hasenfratz and H.\ Leutwyler, Nucl. Phys. B {\bf 343}, 241 (1990).

\bibitem{Leutwyler NRD}
H.\ Leutwyler, Phys. Rev. D {\bf 49}, 3033 (1994).

\bibitem{HofmannSpinWave}
C.\ P.\ Hofmann, Phys. Rev. B {\bf 60}, 388 (1999).

\bibitem{HofmannAF}
C.\ P.\ Hofmann, Phys. Rev. B {\bf 60}, 406 (1999).

\bibitem{Chakravarty Halperin Nelson}
S.\ Chakravarty, B.\ I.\ Halperin and D.\ R.\ Nelson, Phys. Rev. B {\bf 39},
2344 (1989).

\bibitem{Neuberger Ziman}
H.\ Neuberger and T.\ Ziman, Phys.Rev. B {\bf 39}, 2608 (1989).

\bibitem{Fisher}
D.\ S.\ Fisher, Phys. Rev. B {\bf 39}, 11783 (1989). 

\bibitem{Hasenfratz Niedermayer}
P.\ Hasenfratz and F.\ Niedermayer, Z. Phys. B: Condens. Matter {\bf 92}, 91
(1993).

\bibitem{Introductions}
B.\ Borasoy, in {\it The standard model and beyond: Proceedings}, edited by
T.\ Aliev, N.\ K.\ Pak, M.\ Serin (Berlin, Springer, 2008), p.1;
C.\ P.\ Burgess, Ann. Rev. Nucl. Part. Sci. {\bf 57}, 329 (2007); 
V.\ Bernard and U.-G.\ Meissner,  Ann. Rev. Nucl. Part. Sci. {\bf 57}, 33
(2007);
B.\ Kubis, Lectures given at the Workshop on Physics and Astrophysics of
Hadrons and Hadronic Matter, Shantiniketan, India, 2006, hep-ph/0703274
(unpublished);
S.\ Scherer and M.\ R.\ Schindler, hep-ph/0505265 (unpublished); 
J.\ L.\ Goity, Czech. J. Phys. {\bf 51}, B35 (2004);
S.\ Scherer, Adv. Nucl. Phys. {\bf 27}, 277 (2003);
B.\ R.\ Holstein, hep-ph/0010033 (unpublished); hep-ph/9510344 (unpublished);
A.\ Pich, hep-ph/9806303 (unpublished);
V. Koch, Int. J. Mod. Phys. E {\bf 6}, 203 (1997);
G.\ Ecker, hep-ph/9608226 (unpublished); Prog. Part. Nucl. Phys. {\bf 35}, 1
(1995);
A.\ V.\ Manohar, in {\it Schladming 1996: Perturbative and Nonperturbative
Aspects of Quantum Field Theory}, edited by H.\ Latal and W.\ Schweiger
(Springer, New York,1997), p.311.

\bibitem{Outlines}
B.\ Moussallam, Lectures given at 20th General FANTOM Study Week on QCD at Low
Energies, Emmen, the Netherlands, 2004, hep-ph/0407246 (unpublished);
G.\ Colangelo and G.\ Isidori, Lectures given at the 2000 LNF Spring
School, Frascati, hep-ph/0101264 (unpublished);
J.\ Gasser, Nucl. Phys., Proc. Suppl. {\bf 86}, 257 (2000);
C.\ P.\ Burgess, in {\it Radiative Corrections: Application of Quantum Field
Theory to Phenomenology}, Barcelona, 1998, p.471; hep-ph/9812468
(unpublished); 
H.\ Leutwyler, hep-ph/9409422 (unpublished).

\bibitem{HofmannFerro}
C.\ P.\ Hofmann, Phys. Rev. B {\bf 65}, 094430 (2002).

\bibitem{Roman Soto}
J.\ M.\ Rom\'an and J.\ Soto, Int. J. Mod. Phys. B {\bf 13}, 755 (1999); Phys.
Rev. B {\bf 62}, 3300 (2000); Ann. Phys. (N.Y.)  {\bf 273}, 37 (1999).

\bibitem{Burgess}
C.\ P.\ Burgess, Phys. Rept. {\bf 330}, 193 (2000); C.\ P.\ Burgess and
C.\ A.\ Lutken, Phys. Rev. B {\bf 57}, 8642 (1998).

\bibitem{Uwe Construction}
F.\ K\"ampfer, M.\ Moser and U.-J.\ Wiese, Nucl. Phys. B {\bf 729}, 317 (2005);
C.\ Br\"ugger, F.\ K\"ampfer, M.\ Moser, M.\ Pepe and U.-J.\ Wiese, Phys. Rev.
B {\bf 74}, 224432 (2006).

\bibitem{Uwe CEP}
U.\ Gerber, C.\ P.\ Hofmann, F.-J.\ Jiang, M.\ Nyfeler and U.-J.\ Wiese, J.
Stat. Mech.: Theory and Experiment, P03021 (2009).

\bibitem{Uwe Application}
C.\ Br\"ugger, F.\ K\"ampfer, M.\ Pepe and U.-J.\ Wiese, Eur. Phys. J. B
{\bf 53}, 433 (2006);
C.\ Br\"ugger, C.\ P.\ Hofmann, F.\ K\"ampfer, M.\ Pepe and U.-J.\ Wiese,
Phys. Rev. B {\bf 75}, 014421 (2007);
C.\ Br\"ugger, C.\ P.\ Hofmann, F.\ K\"ampfer, M.\ Moser, M.\ Pepe and U.-J.\
Wiese, Phys. Rev. B {\bf 75}, 214405 (2007);
F.-J.\ Jiang, F.\ K\"ampfer, C.\ P.\ Hofmann and U.-J.\ Wiese, Eur. Phys. J. B
{\bf 69}, 473 (2009).

\bibitem{Leutwyler foundations}
H.\ Leutwyler, Ann. Phys. {\bf 235}, 165 (1994).

\bibitem{Weinberg}
S.\ Weinberg, Physica A {\bf 96}, 327 (1979).

\bibitem{footnote2}
Note also, that this counting scheme only applies within a Lorentz-invariant
framework: Loop corrections involving ferromagnetic magnons, e.g., which
follow a quadratic dispersion relation, are suppressed by three powers of
momentum in four space-time dimensions and suppressed by two powers of
momentum in three space-time dimensions.

\bibitem{footnote3}
For a review of the effective Lagrangian method at non-zero temperature, see
refs. \cite{Review finite T}. For a general review of field theory at finite
temperature, see refs. \cite{Landsman Weert Kapusta}.

\bibitem{Review finite T}
H.\ Leutwyler, Nucl. Phys. B (Proc. Suppl.) {\bf 4}, 248 (1988); in
{\it Warsaw International Symposium on Elementary Particle Physics, Kazimierz,
1988 -- New Theories in Physics}, edited by Z. Ajduk, S. Pokorski and A.
Trautman (World Scientific, Singapore, 1989) p. 116; also in {\it Summer
Institute in Theoretical Physics, Kingston, 1988 -- Symmetry Violations in
Subatomic Physics}, edited by B.\ Castel and P.\ J.\ O'Donnell (World
Scientific, Singapore, 1989), p. 57.

\bibitem{Landsman Weert Kapusta}
N.\ P.\ Landsman and C.\ G.\ van Weert, Phys. Rept. {\bf 145}, 141 (1987);
J.\ I.\ Kapusta, {\it Finite-Temperature Field Theory} (Cambridge University
Press, Cambridge, England, 1989); A.\ V.\ Smilga, Phys. Rep. {\bf 291}, 1
(1997); J. Zinn-Justin, hep-ph/0005272 (unpublished).

\bibitem{Gerber Leutwyler}
P.\ Gerber and H.\ Leutwyler, Nucl. Phys. B {\bf 321}, 387 (1989).

\bibitem{Arovas Auerbach}
D.\ P.\ Arovas and A.\ Auerbach, Phys. Rev. B {\bf 38}, 316 (1988); Phys. Rev.
B {\bf 40}, 791 (1989);
A.\ Auerbach and D.\ P.\ Arovas, Phys. Rev. Lett. {\bf 61}, 617 (1988).

\bibitem{Okabe et al}
Y.\ Okabe, M.\ Kikuchi and A.\ D.\ S.\ Nagi, Phys. Rev. Lett. {\bf 61}, 2971
(1988).

\bibitem{Barnes}
T.\ Barnes, Int. J. Mod. Phys. C {\bf 2}, 659 (1991).

\bibitem{Mermin Wagner}
N.\ D.\ Mermin and H.\ Wagner, Phys. Rev. Lett. {\bf 17}, 1133 (1966).

\bibitem{Hasenfratz Niedermayer Mass Gap}
P.\ Hasenfratz and F.\ Niedermayer, Phys. Lett. B {\bf 245}, 529 (1990).

\bibitem{spin stiffness}
U.-J.\ Wiese and H. P.\ Ying, Z. Phys. B {\bf 93}, 147 (1994);
B.\ B.\ Beard, R.\ J.\ Birgeneau, M.\ Greven and U.-J.\ Wiese, Phys. Rev. Lett.
{\bf 80}, 1742 (1998).

\bibitem{epsilonExpansionGL}
J.\ Gasser and H.\ Leutwyler, Phys. Lett B {\bf 184}, 83 (1987); {\bf 188},
477 (1987); Nucl. Phys. B {\bf 307}, 763 (1988).

\bibitem{epsilonExpansionL}
H.\ Leutwyler, Phys. Lett. B {\bf 189}, 197 (1987).

\bibitem{epsilonExpansionG}
M.\ G\"ockeler and H.\ Leutwyler, Phys. Lett. B {\bf 253}, 193 (1991); Nucl.
Phys. B {\bf 350}, 228 (1991).

\bibitem{epsilonExpansionMC}
I.\ Dimitrovic, J.\ Nager, K.\ Jansen and T.\ Neuhaus, Phys. Lett. B
{\bf 268}, 408 (1991).

\end{thebibliography}
\end{document}